\documentclass[12pt,a4paper]{article}
\usepackage{amssymb}
\usepackage{srcltx}
\usepackage{amssymb}
\usepackage{enumerate}
\usepackage{multicol}
\usepackage{epstopdf}
\usepackage{lipsum}
\usepackage{textcomp}
\usepackage{marginnote}
\usepackage{graphicx}
\usepackage{subcaption}
\usepackage[section]{placeins}

\usepackage{amsfonts}
\usepackage{algorithmic}
\usepackage{float}
\ifpdf
  \DeclareGraphicsExtensions{.eps,.pdf,.png,.jpg}
\else
  \DeclareGraphicsExtensions{.eps}
\fi
\usepackage{enumitem}
\setlist[enumerate]{leftmargin=.5in}
\setlist[itemize]{leftmargin=.5in}

\RequirePackage[OT1]{fontenc}
\RequirePackage{amsthm,amsmath, xcolor}
\RequirePackage[numbers]{natbib}
\RequirePackage[colorlinks,citecolor=blue,urlcolor=blue]{hyperref}

 \usepackage[capitalise]{cleveref} 

\newtheorem{theorem}{Theorem}[section]

\newtheorem{example}{Example}

\newtheorem{definition}[theorem]{Definition}
\newtheorem{remark}{Remark}[section]

\newcommand\fdem{$\Box$}

\newcommand\cF{{\cal F}}

\newcommand\cM{{\cal M}}

\newcommand\cV{{\cal V}}

\def\bbr{{\mathbb R}}
\def\bbn{{\mathbb N}}

\def\text#1{\hbox{#1}}
\def\proof{{\noindent \bf Proof. }}
\def\endproof{\mbox{\ $\qed$}}

\def\E{{\bf E}}

\def\P{{\bf P}}

\def\C{{\bf C}}

\def\H{{\bf H}}

\def\u{{\bf u}}

\def\me{\mathrm{e}}

\def\d{\mathrm{d}}
\def\dsl{\displaystyle}
 \def\ri{\right}
\def\lf{\left}
\def\build #1_#2{\mathrel{\mathop{\kern 0pt #1}\limits_{#2}}}

\newcommand{\wh}{\widehat}
\newcommand{\wt}{\widetilde}
\newcommand{\zs}[1]{{\mathchoice{#1}{#1}{\lower.25ex\hbox{$\scriptstyle#1$}}
{\lower0.25ex\hbox{$\scriptscriptstyle#1$}}}}

\numberwithin{equation}{section}
\ifpdf
\hypersetup{
  pdftitle={Logarithmic utility case},
  pdfauthor={S. Albosaily, and S. Pergamenshchikov}
}
\fi

\begin{document}
\title{
Optimal investment and consumption for  Ornstein-Uhlenbeck spread financial markets with logarithmic utility
\thanks{
This work was supported by
 the RSF grant 17-11-01049 (National Research Tomsk State University).
}
}

\author{
Sahar Albosaily
\thanks{
Laboratoire de Math\'ematiques Raphael Salem,
 UMR 6085 CNRS- Universit\'e de Rouen,  France 
 and 
 University of Hail, Saudi Arabia, 
 ORCID iD: 0000-0002-5714-7834,
 e-mail:sahar.albosaily@etu.univ-rouen.fr
}
 and\\
 Serguei Pergamenshchikov\thanks{
 Laboratoire de Math\'ematiques Raphael Salem,
 UMR 6085 CNRS- Universit\'e de Rouen Normandie,  France
and
International Laboratory of Statistics of Stochastic Processes and
Quantitative Finance, National Research Tomsk State University,
 e-mail:
Serge.Pergamenshchikov@univ-rouen.fr } 
}

\date{}

\maketitle

\begin{abstract}
We consider a spread financial market defined by the multidimensional Ornstein--Uhlenbeck (OU) process.
We study the optimal consumption/investment problem for logarithmic utility functions in the base of stochastic dynamical programming method.
We show a special  Verification Theorem for this case.
We find the solution to the Hamilton--Jacobi--Bellman (HJB) equation in explicit form 
and as a consequence we construct the optimal financial strategies.
Moreover, we study the constructed strategy by numerical simulations.
\end{abstract}

{\bf keywords}
 Optimality, 
Feynman--Kac mapping, 
Hamilton--Jacobi--Bellman equation, 
It\^o formula, 
Brownian motion, 
Ornstein--Uhlenbeck process,
Stochastic processes,
Financial market,
Spread market.


 \section{Introduction}
This paper deals with an optimal investment/consumption problem during a fixed time interval $[0,T]$ for a financial market generated by risky spread assets defined through the general multidimensional Ornstein--Uhlenbeck (OU) processes (see, for example, \cite{BoguslavskyBoguslavskaya2004} and \cite{AlbosPerga2017} ). 
The idea of spread market goes back to the $1980$'s where the team of Nunzio Tartaglia   at Morgan Stanley proposed the pairs trading idea to take advantage of market mispricing to gain profit \cite{Elliottet.al.2005}. 
Several studies have been using the notion of spread to examine the behaviour of  financial market. 
For example for the precious metals spread, it has been examined  the spread of gold future market and the Treasury bill future market by \cite{MonroeCohn1986}  .
In addition, it has been studied for oil markets such as  \cite{GrimaPaulson1999} have been investigated the long term price relationship between futures prices of crude oil and heating oil. 
The idea of pairs trading is widely used however the academic research about it is still small \cite{Krauss2017}. 
In this paper we are concerned on the time-series approach of pairs-trading. It is been proposed in \cite{Elliottet.al.2005} the mean-reverting Gaussian Markov chain model and \cite{Reverre2001}  has discussed the classical study of pairs trading of Royal Dutch and Shell stocks. 
Also in other sectors like the microstructioe level within the airline industry (see, for example, \cite{ZebedeeKasch-Haroutounian2009}) as well it is known in many hedge funds \cite{CaldeiraMoura2013}.
\\
Moreover, these problems for Black-Scholes (Bl-Sch) market and stochastic utility market are considered in many papers (see for example \cite{Karatzasshreve1998}, \cite{KluppelbergPergamenchtchikov2009}, \cite{DuffieFilipovicSchachermayer2003} and \cite{BerdjanePergamenchtchikov2013}).
The affine processes proposed in 
 \cite{DuffieFilipovicSchachermayer2003} and \cite{KraftSeiferlingSeifried2017} to be used in the financial market in the general framework, however, unfortunately we can not use these methods due to the additional variable in the HJB equation corresponding to the risky asset.
So in this paper we investigate the optimal investment/consumption  problem for logarithm utility functions with no constraints or transaction fees over the whole investment interval $[0,T]$. 
Using the stochastic dynamical programming method in solving this type of problems, we obtain all optimal solutions in explicit form.
To this end we studied the Hamilton--Jacobi--Bellman (HJB) equation  and we found its solution in an explicit form. 
We shown a special new verification theorem for this case and making use of this theorem we construct the optimal strategy.
The main difference between this model and Black - Scholes model is that in this model we obtained in the HJB equation the additional multivariate spread variables corresponding to the O-U process. 
By these reasons, we need to develop a new analytical tool method for this optimisation problem.

The rest of the paper is organized as follows. 
In section~\ref{sec:Mm}
we formulate the problem and we define the price process for the Ornstein-Uhlenceck model
In section~\ref{sec:HJB} 
we write the HJB equation. 
In section~\ref{sec:MnRslt} 
we state the main results of the paper.
Numerical simulations are given in section \ref{sec:Nsim}.
The corresponding verification theorem is stated in section~\ref{VTh: A1}
Some auxiliary results are stated in the Appendix.



\section{Market model}\label{sec:Mm}
Let 
$ (\Omega, \cF_\zs{T}, (\cF_\zs{t})_\zs{0\le t\le T}, \P)$
be a standard filtered probability space with 
$ (\cF_\zs{t})_\zs{0\le t\le T}$ 
adapted Wiener processes 
$W= (W_\zs{t})_\zs{0\le t\le T} \in \bbr^m$. 
Our financial market consists of one {\em riskless bond}  
$ (\check{S}_\zs{t})_\zs{0\le t\le T}$ 
and {\em risky spread stocks}
$ (S_\zs{t})_\zs{0\le t\le T}$ 
governed by the following equations:
\begin{equation}\label{sec:Mm.1}
\left\{  \begin{array}{ll}
		\d \check{S}_\zs{t}& = r\check{S}_\zs{t} \d  t, 
	         \hfill
		\check{S}_\zs{0} = 1 ,
	        \\[2mm]
                  \d  S_\zs{t}& =  A S_\zs{t} \d  t+ \sigma \d  W_\zs{t},
	          \qquad
		S_\zs{0}>0 \,,
		\end{array}
\right.
\end{equation}
where $r \ge 0$ is the interest rate for riskless asset,  the $d$ vector  risky assets
\\$ S_\zs{t}=(S_\zs{1}(t), S_\zs{2}(t), S_\zs{3}(t), \dots, S_\zs{d}(t)) $, 
 the standard Brownian motion $( W_\zs{t})_\zs{0 \le t \le T}$ with values in 
$ \bbr^m$,  
the volatility 
$ \sigma$ is a $d \times m$ matrix such that $(\sigma \sigma')^{-1}$ exists, 
and the $d \times d$ mean reverting matrix $A$ is given by
\begin{equation} \label{eq:A}
\, A=  \begin{pmatrix} 
\begin{array}{ccccc} 
								a_\zs{11} & a_\zs{12} & \dots & a_\zs{1d} \\
								a_\zs{21} &  a_\zs{22} & \dots & a_\zs{2d} \\
								\vdots & & \ddots &  \\
								a_\zs{d1} & a_\zs{d2} & \dots & a_\zs{dd} 
\end{array}
\end{pmatrix} \,,
\end{equation}
with negative real eigenvalues i.e. 
$\mbox{Re} \lambda_\zs{i} (A)<0$. 
Let now 
$ \check{\alpha}_\zs{t}$ be the number of riskless assets $ \check{S}$ 
and 
$\alpha_\zs{t}=(\alpha_\zs{1}(t), \alpha_\zs{2}(t), \dots, \alpha_\zs{d}(t) ) \in \bbr^d$  be the number of risky assets at the moment $ 0 \le t \le T $,
 and the consumption rate is given by a non negative  integrated function $ (c_\zs{t})_\zs{0\le t\le T}$ \cite{Karatzasshreve1998}.
Thus the wealth process  for 
$X_\zs{t}= \check{\alpha}_\zs{t} \check{S}_\zs{t} +  \alpha'_\zs{t} S_\zs{t}$
is given by
$$
 \d  X_\zs{t} =   \check{\alpha}_\zs{t} \d \check{S}+ \alpha_\zs{t}'  \d  S_\zs{t} -  c_\zs{t}  \d  t \,,
$$
which can be written as
\begin{equation}\label{sec:Md.3}
 \d  X_\zs{t}   =  (r	X_\zs{t} -   \alpha_\zs{t}' \wh{S}  -  c_\zs{t} ) \d  t 
 					+ 	\alpha'_\zs{t} \sigma  \d  W_\zs{t} \,,
\end{equation}
where  $\wh{S}= A_\zs{1} S= (\wh{S}_\zs{1}, \dots, \wh{S}_\zs{d})' \in \bbr^d$ 
and $A_\zs{1}= r I_\zs{d} - A$,  the prime $'$  denotes the transposition. 
Note that in this case the matrix $A_\zs{1}$ is invertible, i.e. there exists $A_\zs{1}^{-1}$.
In this paper we use the logarithmic utility functions, i.e., we need the following definition for the admissible strategies.
\begin{definition}
The strategy $ \upsilon = (\upsilon_\zs{t})_\zs{0 \le t \le T}$ is called admissible if it is adapted,  equation \eqref{sec:Md.3} has a unique positive strong solution and the following conditions hold
$$
\E \Big( \int_\zs{0}^T (\ln c_\zs{t})_\zs{-} \d t \Big) < + \infty 
\quad
\mbox{and}
\quad
\E \sup_\zs{0 \le t \le T} \big( \ln (X_\zs{t}^\upsilon ) \big)_\zs{-} < + \infty \,.
$$
We denote by $\cV$ the set of all admissible strategies.
\end{definition}

\noindent 
Now for any $\upsilon \in \cV$ and $\varsigma = (X, S) \in \bbr^N$, where $N= d+1$, 
we define the objective function as 
$$
\pmb{J}(\varsigma, \upsilon) : = \E_\zs{\varsigma} \Big( \int_\zs{0}^T (\ln c_\zs{u}) \d u + \varpi \ln (X_\zs{T}^\upsilon) \Big) \,,
$$
where $\E_\zs{\varsigma}$ is the expectation under condition $\varsigma_\zs{0}= \varsigma = (x, s)$. 
Our goal in this paper is to maximize this function, i.e.
\begin{align}\label{sec:Model.01++*}
\pmb{J}^*(\varsigma) : =  \sup_{\upsilon \in \mathcal{V}} \pmb{J}(\varsigma,\upsilon)\,.
\end{align}
To study this problem we use the stochastic dynamic programming method. 
To this end we need to study the  value functions $(\pmb{J}^*(\varsigma, t))_\zs{0\le t\le T}$ defined as
$$
\pmb{J}^*(\varsigma, t)=\,
\sup_{\upsilon \in \mathcal{V}}  \E_\zs{\varsigma, t} \Big( \int_\zs{t}^T (\ln c_\zs{u}) \d u + \varpi  \ln (X_\zs{T}^\upsilon) \Big)
 \,,
$$
where $\varpi >0 $ and  $\E_\zs{\varsigma, t}$ is the expectation under condition $\varsigma_\zs{0}=  \varsigma = (x, s)$.
Thus we need to study the HJB equation which is given in the following section.

\section{Hamilton--Jacobi--Bellman equation}\label{sec:HJB}
Denoting by  $ \varsigma_\zs{t} = (X_\zs{t},S_\zs{t})'  \in \bbr^{N}$ where 
\\
$N=d+1$,
we can rewrite  the  wealth and stock equations given in   \cref{sec:Mm.1} and \eqref{sec:Md.3} respectively
in the following form
\begin{equation}\label{sec:HJB.1}
 \d  \varsigma_\zs{t} = \check{a}(\varsigma_\zs{t}, \upsilon_\zs{t})  \d  t + \check{b} (\varsigma_\zs{t}, \upsilon_\zs{t})  \d  W_\zs{t},
 \quad
 \varsigma_\zs{0}= \varsigma \,,
\end{equation} 
where
$\check{a} \in \bbr^N$ and $\check{b}$ is the matrix of $N \times m$ functions such that for any $\varsigma = (x, s) \in \bbr^N$
$$
\check{a}(\varsigma, \u) = \begin{pmatrix} 
\begin{array}{c} r x- \alpha' A_\zs{1}  s	 - c \\
											As
											\end{array}
											\end{pmatrix}
\quad
\mbox{and}
\quad
\check{b}(\varsigma, \u ) =\begin{pmatrix} 
\begin{array}{c}   \alpha' \sigma \\
 \sigma  
   \end{array} \end{pmatrix}  \,,
$$
with the control variable $\u = (\alpha, c)$ with $\alpha \in \bbr^d$ and $c>0$.
Now, for any $\pmb{q} =(\pmb{q}_\zs{1}, \dots, \pmb{q}_\zs{N})'  \in \bbr^N$ and  $N \times N$ symmetric matrix 
$\pmb{M}= (\pmb{M}_\zs{ij})_\zs{1 \le i,j  \le N}$,
 we set the Hamilton function as 
\begin{equation} \label{sec:HJB.2}
 H(\varsigma, \pmb{q}, \pmb{M}): =  \sup_{\u \in \Theta} H_0(\varsigma, \pmb{q}, \pmb{M}, \u), \qquad \Theta \in \bbr^d \times \bbr_\zs{+},
\end{equation}
 where
 $$
  H_0( \varsigma, \pmb{q}, \pmb{M}, \u): =  \check{a}'(\varsigma, \u) \pmb{q}+ \frac{1}{2} \textbf{tr}[\check{b} \check{b}'(\varsigma, \u) \pmb{M}] +\ln c \,.
$$
 \\
In order to study problem  \eqref{sec:Model.01++*}, we need to solve the HJB equation  which is given by
\begin{align} \label{Hamilton--Jacobi--Bellman equation}
\begin{cases}
 z_\zs{t} (\varsigma, t)+H(\varsigma, \partial z(\varsigma, t), \partial^2 z(\varsigma, t)) = 0 , 
 \qquad
  t \in [0,T] ,
  \\[2mm]
 z(\varsigma, T) = \varpi \ln x, 
 \hfill 
 \varsigma \in \mathbb{R}^N ,
\end{cases}
\end{align}
where 
$\partial z(\varsigma, t) =( z_\zs{x}, z_\zs{s_\zs{1}},\ldots, z_\zs{s_\zs{d}})'\in\bbr^{N}$ and
  $$
 \partial^2 z(\varsigma, t) = \begin{pmatrix}
 										\begin{array}{ccccc}
 										z_\zs{xx} & z_\zs{xs_1} & z_\zs{xs_2} & \dots & z_\zs{xs_d}
 										\\
 										 z_\zs{xs_1} & \ z_\zs{s_1s_1} & z_\zs{s_1s_2} & \dots & z_\zs{s_1s_d} 
 										 \\
 										  \vdots & & &\ddots
 										  \\
 										  z_\zs{xs_d} & z_\zs{s_ds_1} & z_\zs{s_ds_2} & \dots  & \ z_\zs{s_ds_d}
 										  \end{array}
 										   \end{pmatrix}_{N \times N} \,.
 										   $$
To calculate the Hamilton function \eqref{sec:HJB.2}, note that
\begin{align*}
H_0(\varsigma, \pmb{q}, \pmb{M}, \upsilon)=& ( r x- \alpha' \wh{s}	 - c ) \pmb{q}_\zs{1}
										 + \sum^{d}_\zs{i=1} \wt{s}_\zs{i} \pmb{q}_\zs{1+i} 
										\\
										&+ \frac{1}{2} \Big( \alpha'  \sigma \sigma' \alpha  \pmb{M}_\zs{11}
										+ 2 \sum^d_\zs{i=1} < \sigma \sigma' \alpha>_\zs{i}  \pmb{M}_\zs{1, 1+i } 
										\\
										&+ \sum^d_\zs{k,i=1}  <\sigma \sigma' >_\zs{ki} \pmb{M}_\zs{1+k, 1+i } \Big) +  \ln c \,,
\end{align*}
where $\wt{s}= A s =(\wt{s}_\zs{1}, \dots , \wt{s}_\zs{d})' \in \bbr^d$.  
The symbol $<X>_\zs{i}$ denotes the 
$i^{\mbox{\begin{scriptsize}
th
\end{scriptsize}}}$ element of the vector $X$
and $<Y>_\zs{ij}$ denotes the 
$(i,j)^{\mbox{\begin{scriptsize}
th
\end{scriptsize}}}$ element of the matrix $Y$.
Note that due to \eqref{sec:HJB.2}, 
if $\pmb{M}_\zs{11} \ge 0$ or $ \pmb{q}_\zs{1} \le 0$ then the Hamilton function $H(\varsigma, \pmb{q}, \pmb{M}) =+ \infty$.
 So,  we maximize the function $H_0(\varsigma, \pmb{q}, \pmb{M}, \upsilon)$ over $\alpha$ and $c$
 under condition that $\pmb{M}_\zs{11} < 0$ and $\pmb{q}_\zs{1} > 0$. We obtain that optimal values for this maximization problem are given by
\begin{equation}\label{eq: alpha,c}
\alpha^0(s,\pmb{q},\pmb{M})= \frac{ (\sigma \sigma')^{-1} \tau}{ \pmb{M}_\zs{11}} 
\quad
\mbox{and}
\quad
c^0(s,\pmb{q},\pmb{M})=	\frac{1}{\pmb{q}_\zs{1}} 	 \,,
\end{equation}
where $\tau=  \pmb{q}_\zs{1} \wh{s} - \sigma \sigma' \mu  $ and $\mu= (\pmb{M}_\zs{1, 1+1}, \dots, \pmb{M}_\zs{1, 1+d})' $. 
 Now we replace $\alpha_\zs{i}^0$ and $c^0$  into $H_\zs{0}$ to obtain the Hamilton function, so we get
\begin{align*}
H(\varsigma, \pmb{q}, \pmb{M}) = & r x \pmb{q}_\zs{1} 
													-\ln \pmb{q}_\zs{1}
													+\frac{ \tau' (\sigma \sigma')^{-1} \tau}{2 |\pmb{M}_\zs{11}|}
													+ \sum_\zs{i=1}^d \wt{s}_\zs{i} \pmb{q}_\zs{1+i}
													\\
													&+ \sum_\zs{k, i=1}^d <\sigma \sigma'>_\zs{ki} \pmb{M}_\zs{1+i, 1+k}
													-1
												     \,.
\end{align*}

From the preceding Hamilton function and  the HJB equation \eqref{Hamilton--Jacobi--Bellman equation}, we obtain 
 \begin{align} 
z_\zs{t}  +r x z_\zs{x}
			+\frac{ \tau' (\sigma \sigma')^{-1} \tau}{2 |z_\zs{xx}|} 
			-1
			-\ln z_\zs{x}
			+ \sum_\zs{i=1}^d \wt{s}_\zs{i} z_\zs{s_\zs{i}}
             \label{sec:HJB.4}
             + \sum_\zs{k, i=1}^d <\sigma \sigma'>_\zs{ki} z_\zs{s_\zs{i} s_\zs{k}}=0 \,,
 \end{align}
 where $ z(\varsigma, T) = \ln x $ 
 for any $\varsigma \in  \bbr_\zs{+} \times \bbr^d $. 
 To write the solution for this equation, we need to introduce the
   $d \times d$ matrix
 $g= (g_\zs{ij} )_\zs{1 \le i,j \le d}$ which is the solution of the following differentiable equation
\begin{equation} \label{eq: gmatrix}
\dot g + \frac{1}{2} \rho(t) A_\zs{1}' (\sigma \sigma')^{-1} A_\zs{1} - A' (g+ g') =0,
 \quad
 g(T)=0 \,.
\end{equation}
Here, the dot $" \cdot "$ denotes the  derivative. 
Moreover, we set
\begin{equation} \label{eq: f}
 f(t)=  \sum^d_\zs{k,i=1} <\sigma \sigma' >_\zs{ki}( \wt g_\zs{ki}(v) + \wt g_\zs{ik}(v)) +f_\zs{0}(t) \,,
\end{equation}
where
$\wt g(t)=  \int^T_\zs{t}  g(v) \d v$,
$$
f_\zs{0}(t)= \frac{1}{2} r \Big( t^2- 2t(T+1) +T(T+2)  \Big) + \rho(t) \ln \rho(t) 
\quad\mbox{and}\quad
\quad \rho(t)= T-t+1\,.
$$
 We show that 
the  equation \eqref{sec:HJB.4} has the following solution
 \begin{equation} \label{eq: z}
z(x, s, t)= \rho(t) \ln x + s' g(t) s +f(t)\,.
\end{equation}

\begin{remark}
As we see in the HJB equation, the additional variable $s \in \bbr^d$  is the main difference from the Bl-Sch market.
\end{remark}

\section{Main results}\label{sec:MnRslt}
First of all we have to study the HJB equation \eqref{sec:HJB.4} to calculate the value function \eqref{sec:Model.01++*}.

\begin{theorem}\label{thm:solHJB}
The function \eqref{eq: z} satisfies the HJB equation \eqref{Hamilton--Jacobi--Bellman equation}.
\end{theorem}

\noindent Furthermore, to construct the optimal strategies we set
$$
\check{\alpha} \big( \varsigma, t \big) = \alpha^0(\varsigma, \partial z, \partial^2 z)
		=  - (\sigma \sigma')^{-1} \wh{s} x	
\quad
\mbox{and}
\quad
\check{c}(\varsigma, t)	= c^0(\varsigma, \partial z, \partial^2 z)=  \frac{ x}{\rho(t)} \,.
$$
Recall that $\wh{s}= A_\zs{1} s= (\wh{s}_\zs{1}, \dots, \wh{s}_\zs{d})' \in \bbr^d$ .
Using these functions
we define the optimal strategies  $\upsilon^*=(\alpha^*, c^*)$ 
as
 \begin{align} \label{opt. stgy.11}
\alpha^*(t)  
 =   
 \check{\alpha}(\varsigma^*_\zs{t}, t) 
 =
(\sigma \sigma')^{-1} \wh{S}_\zs{t} X_\zs{t}^* 	
\quad 
\mbox{and} 
\quad
c^*(t) 
 =   
\check{c}(\varsigma^*_\zs{t}, t) 
=   \frac{ X_\zs{t}^*}{\rho(t)}\,.
\end{align}
Here $ \varsigma^*_\zs{t}= (X^*_\zs{t}, S_\zs{t})$ and $X^*_\zs{t}$ is the optimal wealth process defined 
by the following stochastic differential equation
\begin{equation}\label{eq:dXt*}
\d X^*_\zs{t}= X^*_\zs{t} a^*(t)    \d t 
                       + X^*_\zs{t} (b^*(t))'   \d W_\zs{t} ,
\quad
X_\zs{0}^*= x \,,
\end{equation}
where 
$$
\quad  a^*(t)=r -    \wh{s}'_\zs{t} (\sigma \sigma')^{-1}  \wh{s}_\zs{t} 	 - \frac{1}{ \rho(t)}
\quad
\mbox{and}
\quad
b^*(t)=  \sigma^{-1}  \wh{s}_\zs{t}  		 \,.
$$

\noindent
Now we show that these processes are optimal solutions for the problem \eqref{sec:Model.01++*}.

\begin{theorem}\label{thm:strgy}
The processes \eqref{opt. stgy.11} and \eqref{eq:dXt*} are the optimal strategies for the problem \eqref{sec:Model.01++*} and 
\begin{equation}
J^*(x, s, t)=  z(x, s, t)= \rho(t) \ln x + s' g(t) s +f(t) \,,
\end{equation}
 where $\rho, g$ and $f$ are given in \eqref{eq: gmatrix}.
\end{theorem}
%
%
\begin{example}
For one dimensional case where a riskless  and risky assets are given respectively by
\begin{equation}\label{sec:Mm.3}
\left\{
		\begin{array}{ll}
								\d \check{S}_\zs{t}& = r\check{S}_\zs{t} \d  t, 
			\hfill
								\check{S}_\zs{0} = 1 ,
			\\[2mm]
							 \d  S_\zs{t}& =  -\kappa S_\zs{t} \d  t+ \sigma \d  W_\zs{t},
			\qquad
								S_\zs{0}>0 \,,
		\end{array}
\right.
\end{equation}
where $r \ge 0$ is the interest rate of the riskless asset, $\kappa>0$ and $\sigma$ are respectively the mean reverting speed and the volatility for  risky assets. 
Therefore, for $\kappa_\zs{1}= \kappa +r >0$, the optimal strategies and the HJB equation are given by 
\begin{align*} 
\alpha^*(t)  
 =   
 \check{\alpha}^0(\varsigma^*_\zs{t}, t) 
 =
- \frac{ \kappa_\zs{1} S_\zs{t}  X_\zs{t}^* }{\sigma^2 } 		
\quad 
\mbox{and} 
\quad
c^*(t) 
 =   
\check{c}^{0}(\varsigma^*_\zs{t}, t) 
=   \frac{ X_\zs{t}^*}{\rho(t)}  \,.
\end{align*}
Moreover, the differential wealth process for this example is given by
$$
\d X^*_\zs{t}=X^*_\zs{t}  a^*(t)   \d t    +   X^*_\zs{t}  b^*(t)    \d W_\zs{t} \,,
$$
where $$
\quad  a^*(t)= \check{a}(S_\zs{t}, t)  = r + \kappa_\zs{1}^2 S_\zs{t}^2 / \sigma^2- 1/ \rho(t)
\quad
\mbox{and}
\quad
b^*(t)=  \check{b}(S_\zs{t}, t) =  \kappa_\zs{1} S_\zs{t} / \sigma \,.
$$
\end{example}
%
%
\begin{example}
For multidimensional case where the market assets are given by 
\begin{equation}\label{sec:Mm.2}
\left\{
		\begin{array}{ll}
								\d \check{S}_\zs{t}& = r\check{S}_\zs{t} \d  t, 
			\hfill
								\check{S}_\zs{0} = 1 ,
			\\[2mm]
							 \d  S_\zs{t}& =  A S_\zs{t} \d  t+ \sigma \d  W_\zs{t},
			\qquad
								S_\zs{0}>0 \,,
		\end{array}
\right.
\end{equation}
where $r$ is the interest rate for riskless asset $\check{S}$,  $S_\zs{t}$ is a  $d-$dimensional vector of  risky assets
$ S_\zs{t}=(S_\zs{1}(t), S_\zs{2}(t), S_\zs{3}(t), \dots, S_\zs{d}(t)) \in \bbr^d$, 
$( W_\zs{t})$ is a standard Brownian motion with values in 
$ \bbr^d$,  
the market volatility matrix
$ \sigma= \mbox{diag}(\sigma_\zs{1}, \sigma_\zs{2}, \dots, \sigma_\zs{d})  $, 
and the mean reverting matrix $A$ is given by
$$
\, A=  \begin{pmatrix} 
\begin{array}{ccccc} 
								a_\zs{11} & a_\zs{12} & \dots & a_\zs{1d} \\
								a_\zs{21} &  a_\zs{22} & \dots & a_\zs{2d} \\
								\vdots & & \ddots &  \\
								a_\zs{d1} & a_\zs{d2} & \dots & a_\zs{dd} \,,
\end{array}
\end{pmatrix} \,,
$$
with negative real eigenvalues i.e. 
$\mbox{Re} \lambda_\zs{i} (A)<0$. 
The optimal wealth process $(X_\zs{t}^*)_\zs{0 \le t \le T} $ is defined by the following stochastic equation
\begin{equation*} 
\d X^*_\zs{t}= X^*_\zs{t} a^*(t)   \d t 
                       + X^*_\zs{t} (b^*(t))'    \d W_\zs{t} ,
\quad
X_\zs{0}^*= x \,,
\end{equation*}
where 
$$
\quad  a^*(t)=r +  \sum_\zs{i=1}^d \frac{\wh{S}^2_\zs{i}(t)  }{\sigma^2_\zs{i} } 		 - \frac{1}{ \rho(t)} \,,
\quad
b^*(t)= (b^*_\zs{1}(t), \dots, b^*_\zs{d}(t))' 
\quad
\mbox{and}
\quad 
b^*_\zs{i} (t) = \frac{\wh{S}_\zs{i}(t)  }{ \sigma_\zs{i}} \,.  
$$
Using the preceding  stochastic differential equation,  the optimal strategies \\
 $\upsilon^*=(\alpha^*, c^*)$ for all $0 \le t \le T$ is of the form:
 \begin{align} \label{opt. stgy.11++}
\alpha_\zs{i}^*(t)  
 =   
 \check{\alpha}^0_\zs{i}(\varsigma^*_\zs{t}, t) 
 =
- \frac{\wh{S}_\zs{i}(t)  X_\zs{t}^* }{\sigma^2_\zs{i} } 		
\quad 
\mbox{and} 
\quad
c^*(t) 
 =   
\check{c}^{0}(\varsigma^*_\zs{t}, t) 
=   \frac{ X_\zs{t}^*}{\rho(t)}  \,,
\end{align}
where
$ \check{\alpha}_\zs{t}$ is the number of riskless assets $ \check{S}$ 
and 
$\alpha_\zs{t}=(\alpha_\zs{1}(t), \alpha_\zs{2}(t), \dots, \alpha_\zs{d}(t) ) \in \bbr^d$  be the number of risky assets $S$ at the moment $ 0 \le t \le T $.
\end{example}
\begin{remark}
It should be noted that the behaviour of theses optimal  strategies are described by the transformed spread process $\wh{S}_\zs{t} = A_\zs{1} S'_\zs{t}$. 
In the scalar case this is the same as $S_\zs{t}$.
However, in the general multidimensional case we need to take into account all components of the spread processes.
\end{remark}
\bigskip
\section{Numerical Simulation}\label{sec:Nsim}
\begin{figure}[!h]
\centering
  \includegraphics[width=5cm]{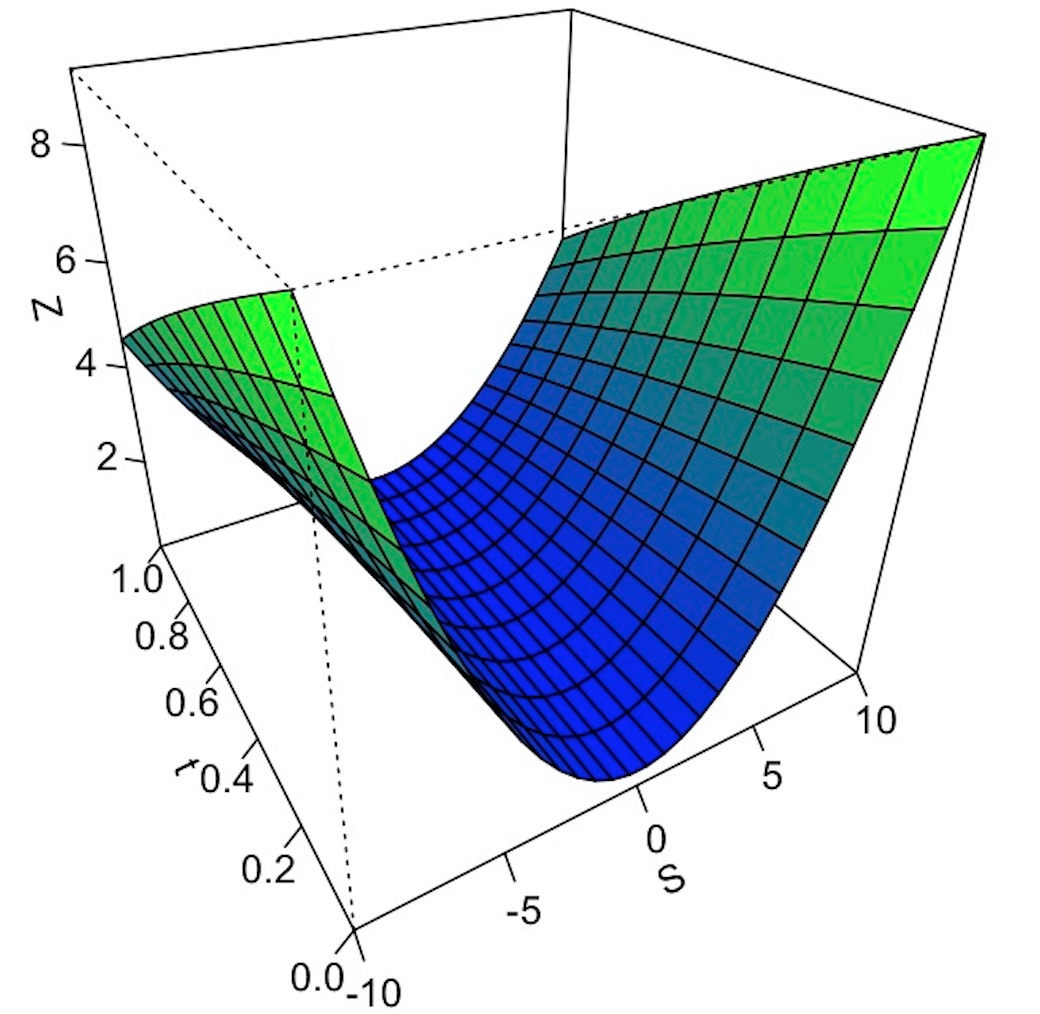}
  \centering
  \caption{\label{fig:boat1}The value function.}
 \end{figure}

\begin{figure}[!h] 
\begin{minipage}[t]{.98\linewidth}
  \begin{subfigure}{0.3\linewidth}
  \includegraphics[scale=0.12]{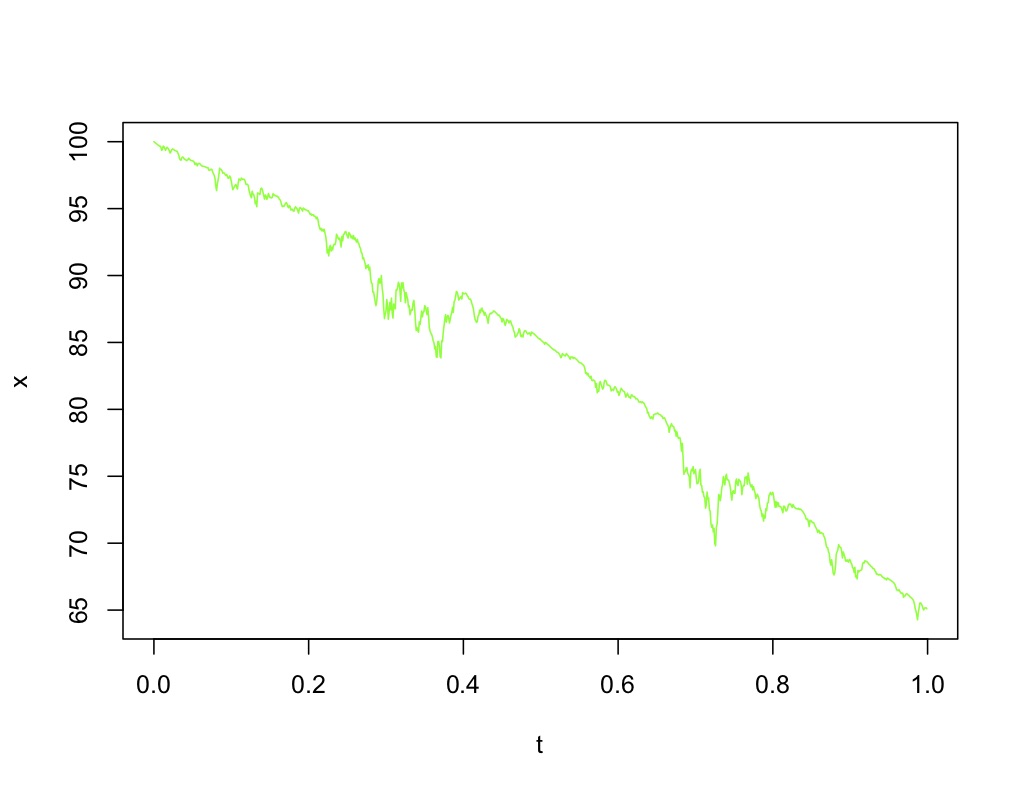}
  \caption{ \label{fig:2.x} \scriptsize{The wealth process $X^*_\zs{t}$.}}
  \end{subfigure}
  \begin{subfigure}{0.3\linewidth}
  \includegraphics[scale=0.12]{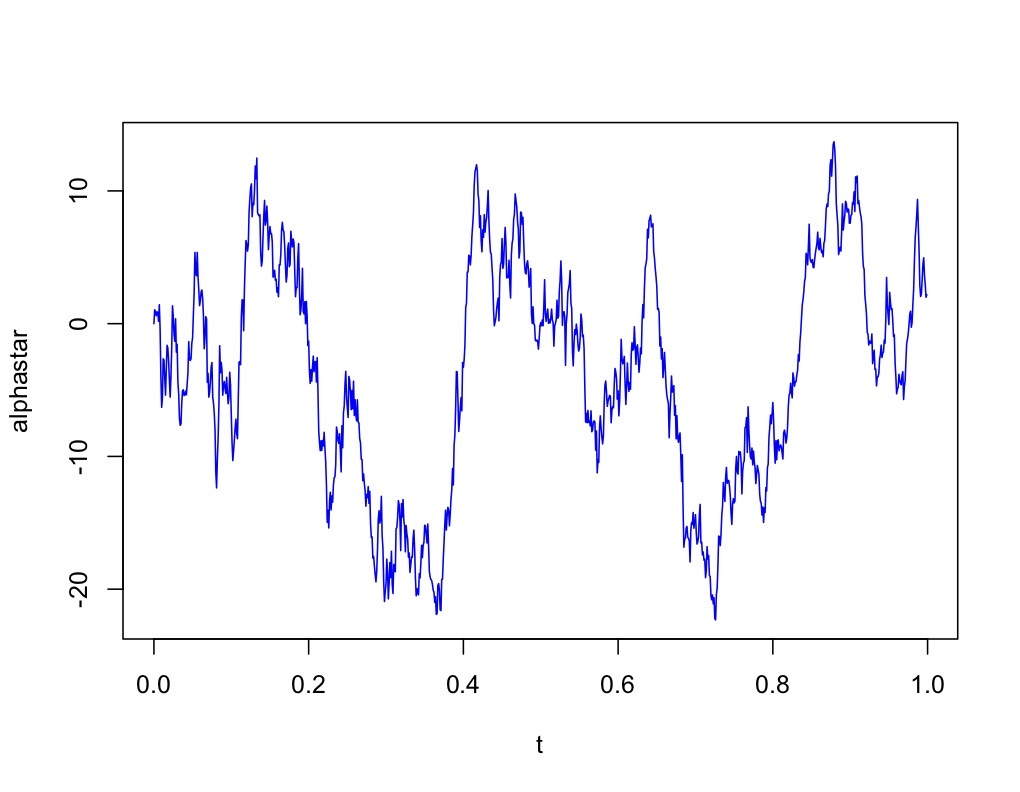}
  \caption{\label{fig:2.alpha} \scriptsize{Optimal investment $\alpha^*_\zs{t}$.}}
  \end{subfigure}
  \begin{subfigure}{0.3\linewidth}
  \includegraphics[scale=0.12]{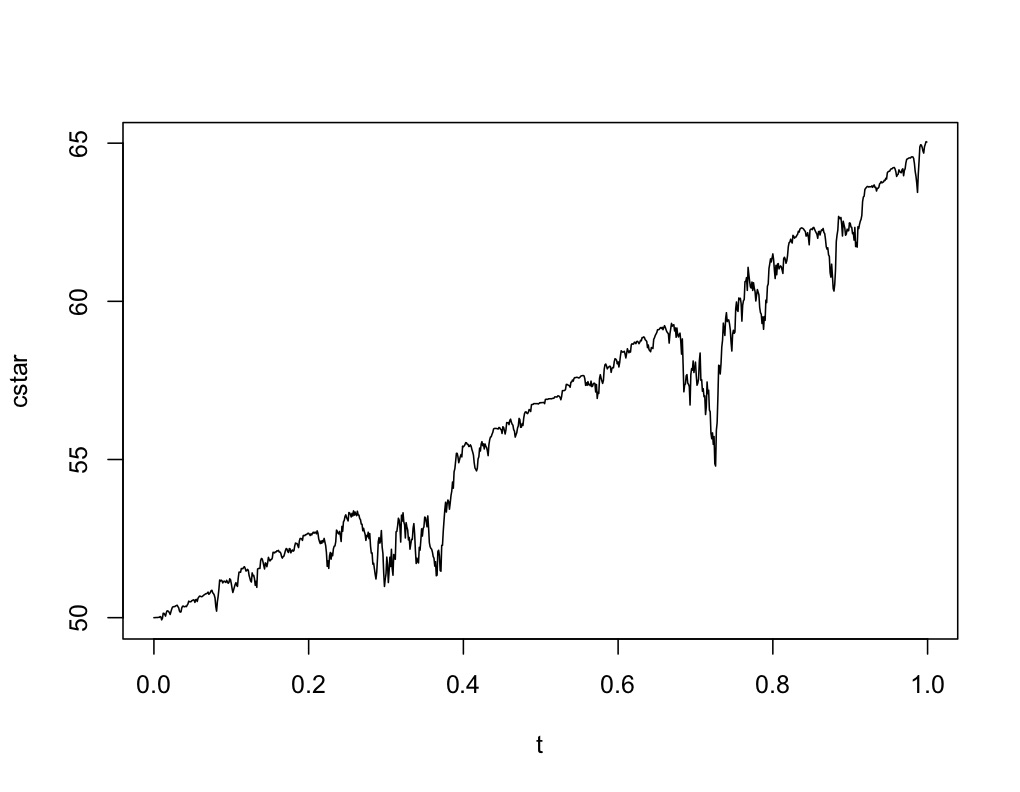}
  \caption{\scriptsize{Optimal consumption $c^*_\zs{t}$.}}
  \end{subfigure}
  \caption{  \label{fig:wealthspread2} The wealth process with the parameters $\alpha$ and $c$ when $\sigma=1$, $r=0.01$ and $\kappa=0.5$.}
  \end{minipage} 
  \end{figure}

\begin{figure}[!h]  
\begin{minipage}[t]{.98\linewidth}
  \begin{subfigure}{0.3\linewidth}
  \includegraphics[scale=0.12]{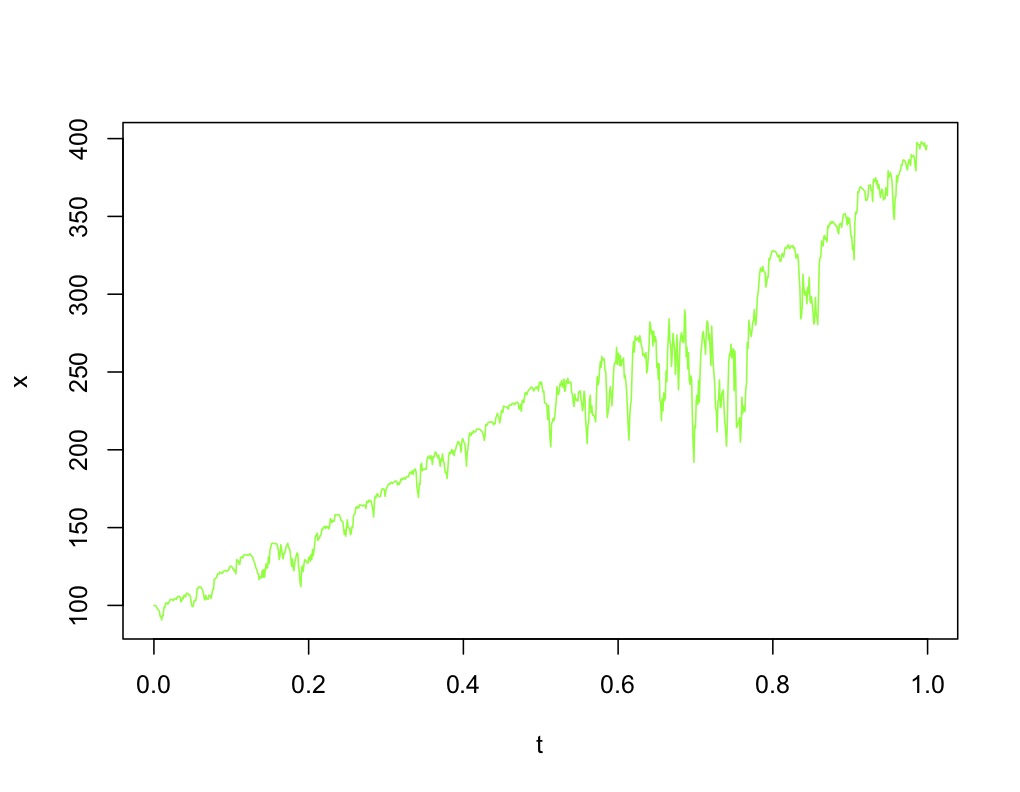}
  \caption{\label{fig:3.x} \scriptsize{The wealth process $X^*_\zs{t}$.}}
  \end{subfigure}
  \begin{subfigure}{0.3\linewidth}
  \includegraphics[scale=0.12]{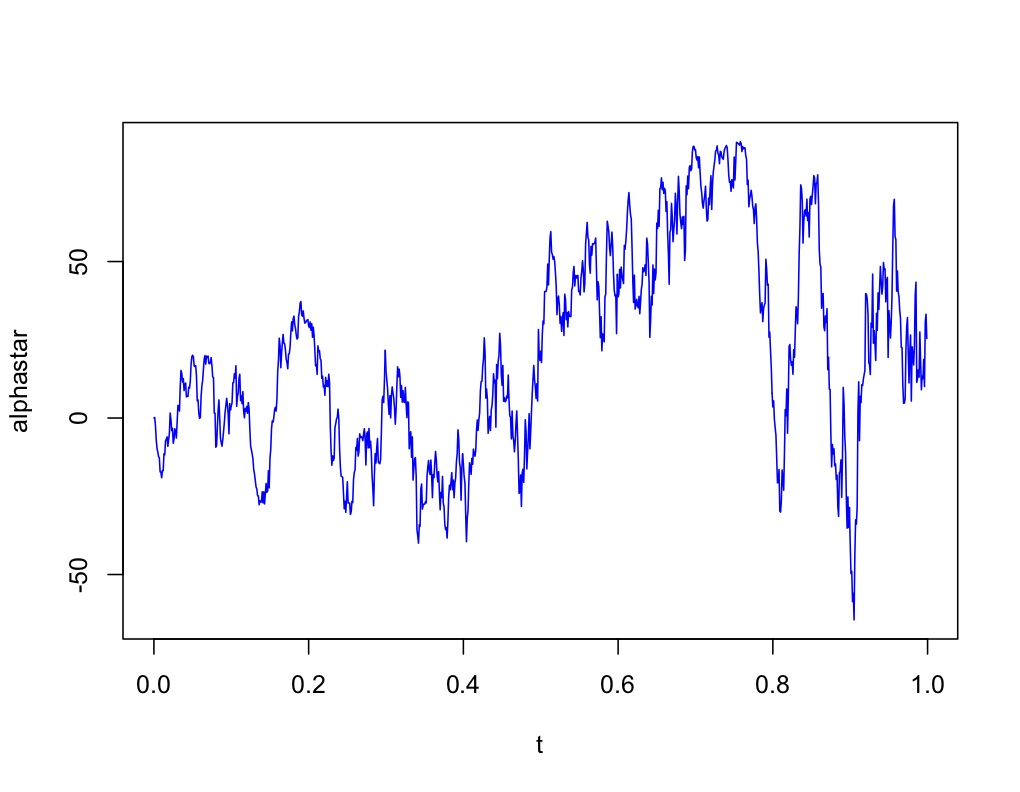}
  \caption{\label{fig:3.alpha} \scriptsize{Optimal investment $\alpha^*_\zs{t}$.}}
  \end{subfigure}
  \begin{subfigure}{0.3\linewidth}
  \includegraphics[scale=0.12]{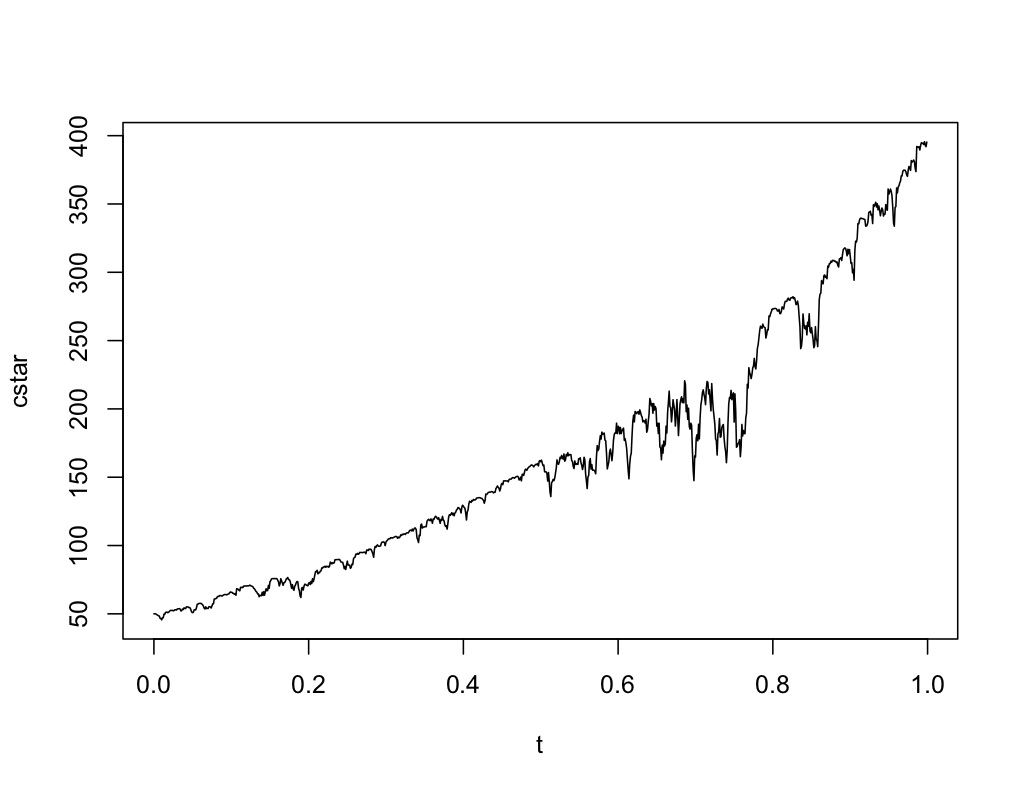}
  \caption{\scriptsize{Optimal consumption $c^*_\zs{t}$.}}
  \end{subfigure}
  \caption{The wealth process with the parameters $\alpha$ and $c$ when $\sigma=5$, $r=4$ and $\kappa=5$.}
   \label{fig:wealthspread3}
  \end{minipage}
  \end{figure}

\begin{figure}[!h]  
\begin{minipage}[t]{.98\linewidth}
\begin{subfigure}{0.3\linewidth}
\includegraphics[scale=0.12]{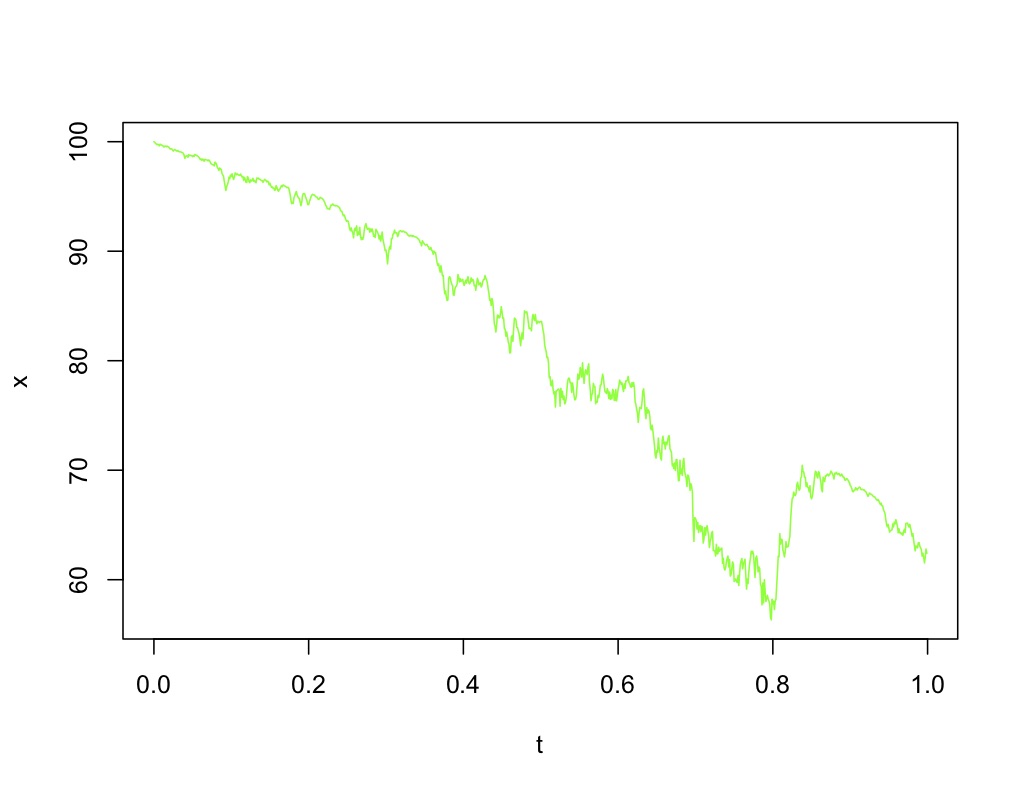}
\caption{\label{fig:4.x}  \scriptsize{Wealth process $X^*_\zs{t}$}}
\end{subfigure}
\begin{subfigure}{0.3\linewidth}
\includegraphics[scale=0.12]{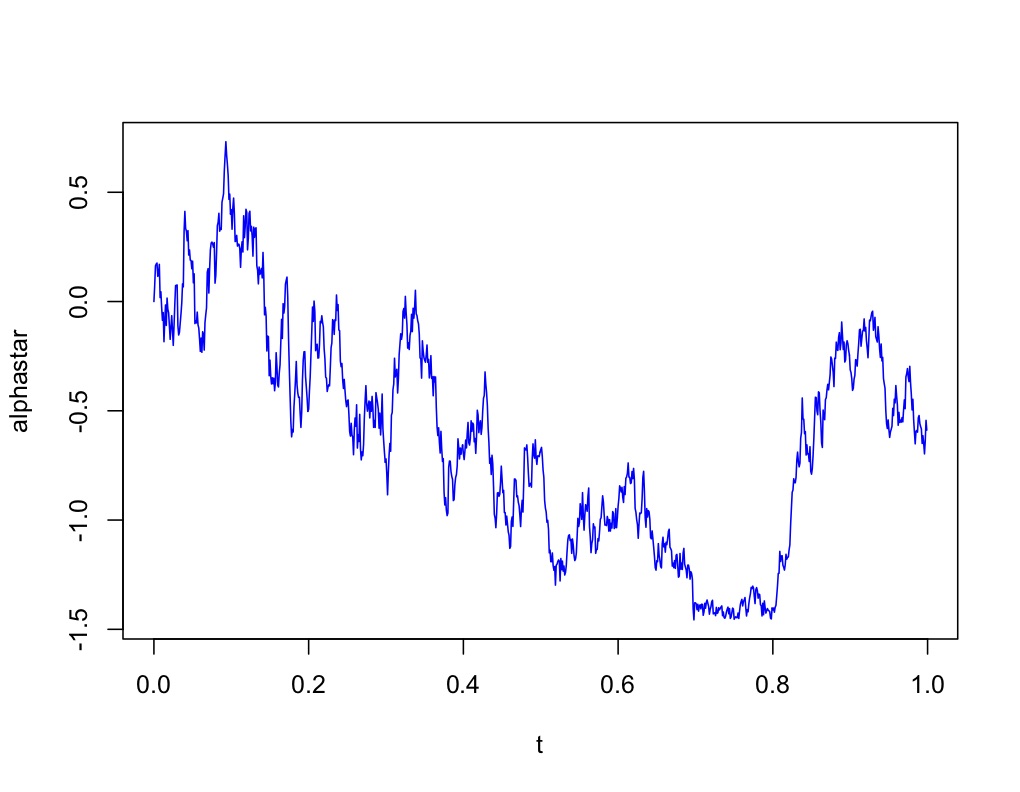}
 \caption{ \label{fig:4.alpha} \scriptsize{Optimal investment $\alpha^*_\zs{t}$.}}
 \end{subfigure}
 \begin{subfigure}{0.3\linewidth}
 \includegraphics[scale=0.12]{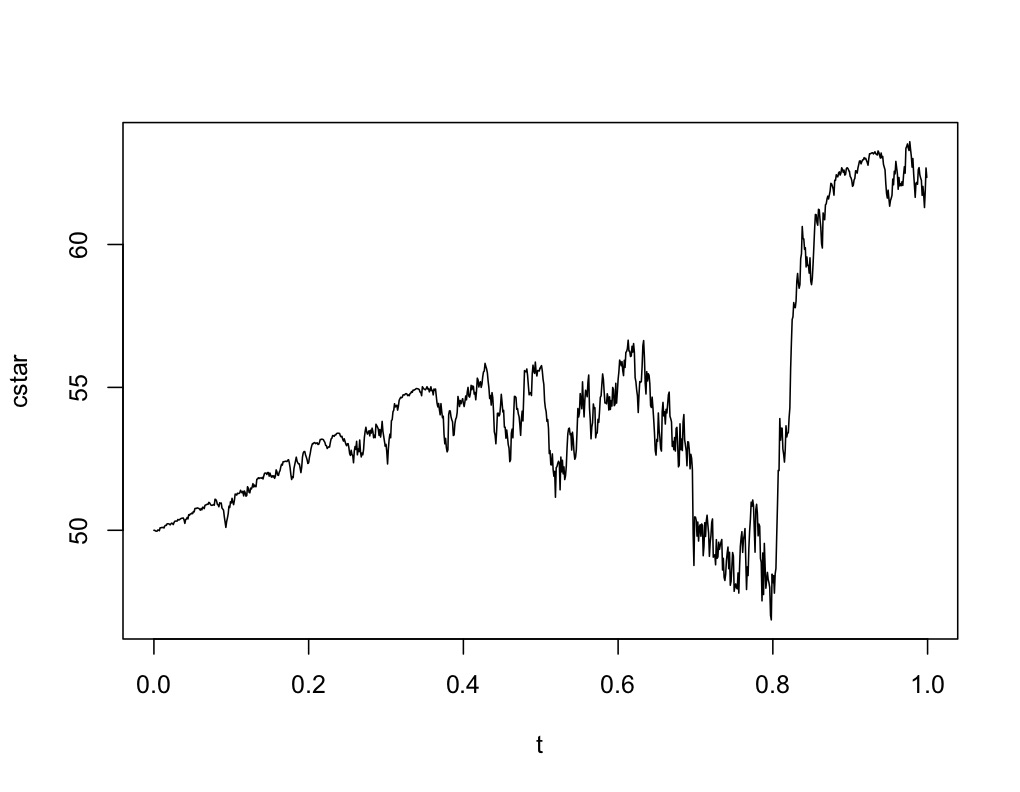}
  \caption{\label{fig:3} \scriptsize{Optimal consumption $c^*_\zs{t}$.}}
 \end{subfigure}
  \caption{The wealth process with the parameters $\alpha$ and $c$ when $\sigma=20$, $r=0.01$ and $\kappa=0.5$ with $n=1000$.}
  \label{fig:wealthspread4}
  \end{minipage}
  \end{figure}

\begin{figure}[!h]  
\begin{minipage}[t]{.98\linewidth}
 \begin{subfigure}{0.3\linewidth}
\includegraphics[scale=0.12]{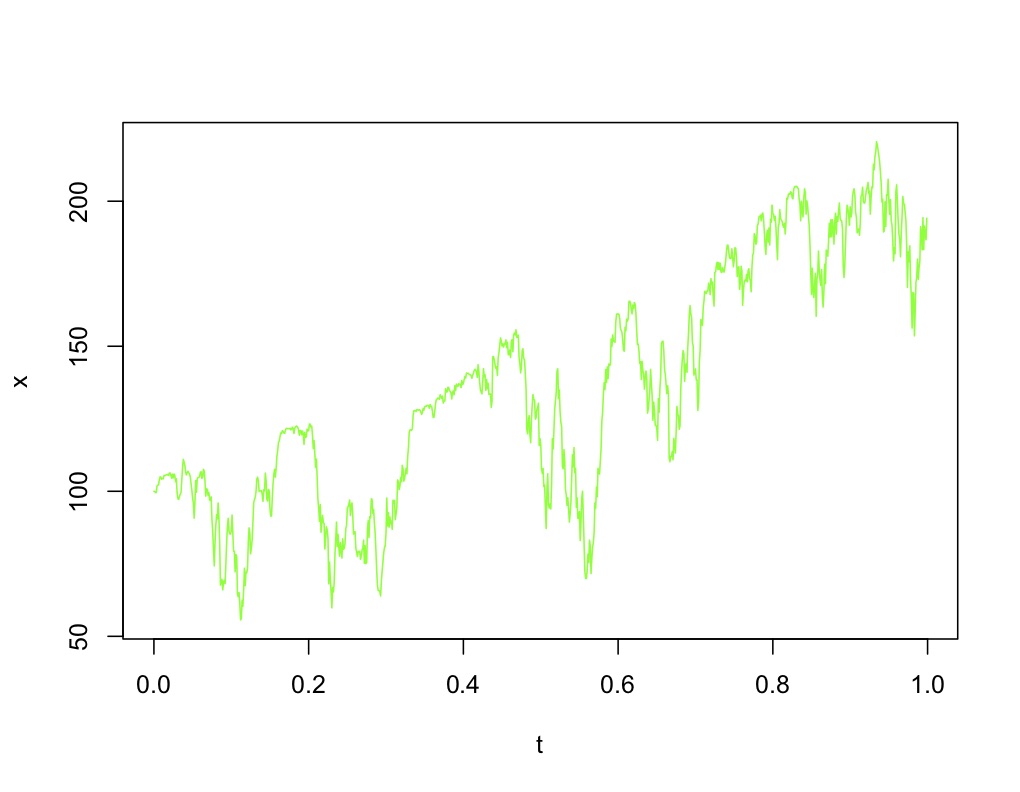}
\caption{\label{fig:5.x}  \scriptsize{The wealth process $X^*_\zs{t}$.}}
 \end{subfigure}
  \begin{subfigure}{0.3\linewidth}
  \includegraphics[scale=0.12]{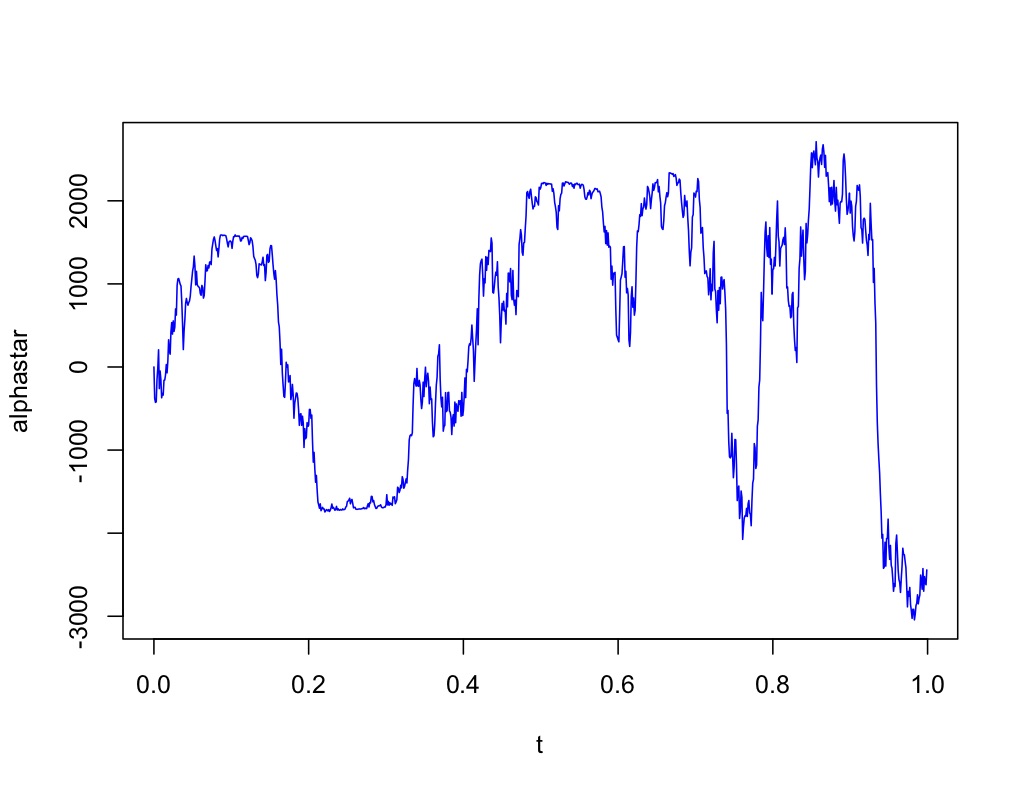}
 \caption{\label{fig:5.alpha}  \scriptsize{Optimal investment $\alpha^*_\zs{t}$.}}
  \end{subfigure}
 \begin{subfigure}{0.3\linewidth}
  \includegraphics[scale=0.12]{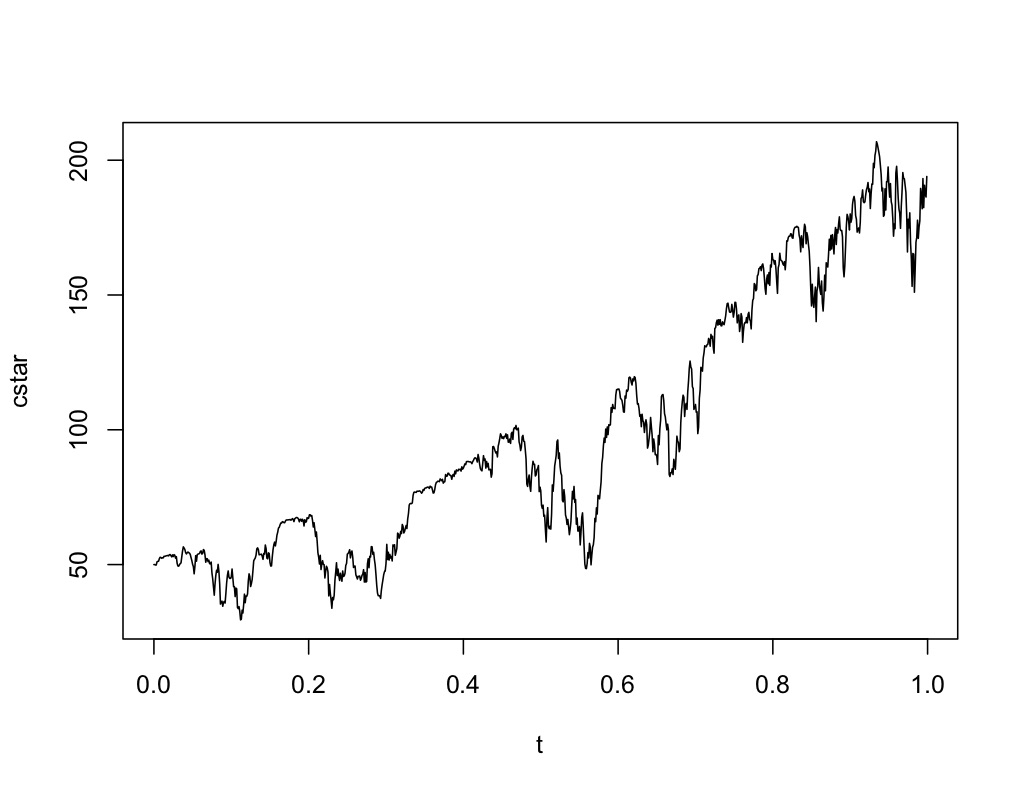}
\caption{\label{fig:a3}  \scriptsize{Optimal consumption $c^*_\zs{t}$.}}
 \end{subfigure}
 \caption{The wealth process with the parameters $\alpha$ and $c$ when $\sigma=0.1$, $r=0.01$ and $\kappa=5$ with $n=1000$.}
  \label{fig:wealthspread5}
\end{minipage}
\end{figure}

%
For $1-$dimensional case. \cref{fig:boat1} shows the value function $z(\varsigma, t)$ given by \cref{eq: z}. 
The following parameters have been used: $T=1$, $r=0.01$, $\kappa=0.1$, $\sigma=0.5$ and the initial endowment $x=100$.
 \\
 Now, we simulate the optimal strategies $\alpha^*_\zs{t}$ and $c^*_\zs{t}$ given in \cref{opt. stgy.11} with the optimal wealth process $x^*_\zs{t}$. 
In the following figures, we used different parameters to show the behaviour of the strategies with different values of $r$,$ \kappa$ and $\sigma$. 
As seen in the figures below, we see that the behaviour of the wealth process is increasing constantly when $\kappa$ has large values (see \cref{fig:3.x} and  \cref{fig:5.x} ). 
However, it is clear that the wealth process is decreasing when $\kappa$ has a quite small value as seen in  \cref{fig:2.x} and \cref{fig:4.x}. 
In addition we see that the volatility in the investment process increases and decreases depending on the fraction $\kappa_\zs{1}/ \sigma^2$.  
Thus the range of volatility in figures (\cref{fig:2.alpha}, \cref{fig:3.alpha} and \cref{fig:4.alpha}) is less than \cref{fig:5.alpha} which jumps to $4000$ points. 
This is due to the higher number we get from the fraction which is nearly $50$.
\\
\bigskip

\section{Verification theorem} \label{VTh: A1}

Now we give some modifications for the verification theorem from \cite{BerdjanePergamenchtchikov2013}.
Consider on the interval $ [0,T] $, the stochastic control process given by  $ N-$dimensional It\^o process
\begin{align}\label{sec:Model.00}
\begin{cases}
d \varsigma^{\upsilon}_\zs{t} & =  \check{a}(\varsigma^{\upsilon}_\zs{t}, t,  \upsilon) dt+ \check{b}(t, \varsigma^{\upsilon}_\zs{t}, \upsilon) dW_\zs{t} ,
 \quad 
 t\geq 0 , 
\\
\varsigma^{\upsilon}_0 & =  x\in \mathbb{R}^N , 
\\
\end{cases}
\end{align}
where $ (W_\zs{t})_\zs{0 \leq  t  \leq  T} $  is a standard $ m-$dimensional Brownian motion. We assume that the control process $ \upsilon$ takes values in some set $ \Theta $. Moreover, we assume that the coefficients $ \check{a}$ and 
$ \check{b}$ satisfy the following conditions:
\begin{itemize}
\item[$V_\zs{1}$)] {\sl For all $ t \in [0,T] $,  the functions $ \check{a}(., t, .)$  and $ \check{b}(., t, .,) $ are continuous on $ \bbr^N \times \Theta ;$} where $ \Theta \in \mathbb{R} \times \mathbb{R}_{+} $. 
\item[$V_\zs{2}$)] {\sl For every deterministic vector $ \upsilon \in \Theta $,  the stochastic differential equation
\begin{equation}
d \varsigma^{\upsilon}_\zs{t}  =  \check{a}(\varsigma^{\upsilon}_\zs{t}, t,  \upsilon) \d t+ \check{b}(\varsigma^{\upsilon}_\zs{t}, t, \upsilon) \d W_\zs{t} \,,
\end{equation}
 with an $N \times m$ matrix $\check{b}$, has a unique strong solution.
}

\end{itemize}

Now we introduce an admissible control process for  equation \cref{sec:Model.00}.
\\
\begin{definition}\label{def:1}
We set 
$$
\mathcal{F}_\zs{t} =  \sigma \{ W_\zs{u}, 0  \leq  u  \leq  t\} ,
\quad
\mbox{for any}
\quad
0 <t  \leq  T ,
$$
where a stochastic control process 
$ \upsilon = (\upsilon_\zs{t})_\zs{t\geq0} = (\alpha_\zs{t}, c_\zs{t})_\zs{t \geq 0}$ 
 is called admissible on $ [0,T] $  
 with respect to equation \cref{sec:Model.00}
  if it is $ (\mathcal{F}_\zs{t})_{0  \leq  t  \leq  T} $  progressively measurable
   with values in $ \Theta$, 
   and equation \cref{sec:Model.00} has a unique strong a.s. continuous solution
    $ (\varsigma_\zs{t}^{\upsilon})_{0  \leq  t  \leq  T} $  such that 
    \begin{equation}
    \E \int^T_\zs{0} \left( \pmb{f} (\varsigma_\zs{u}, u, \upsilon_\zs{u}) \right)_\zs{-}  \d t < + \infty \,,
    \quad
    \E \sup_\zs{0 \le t \le T} ( \pmb{h}( \varsigma^\upsilon_\zs{T}) )_\zs{-} < + \infty \,,
    \end{equation}
    and
\begin{equation}\label{eq: inta+b}
\int_0^T (| \check{a}(\varsigma_\zs{u}^\upsilon, u, \upsilon_\zs{u})|+| \check{b}(\varsigma_\zs{u}^\upsilon, u, \upsilon_\zs{u})|^2 )dt 
				+  \int_\zs{0}^T | \pmb{f} (\varsigma_\zs{u}, u, \upsilon_\zs{u}) | \d u 
				< \infty \quad \mbox{a.s.} \,.
\end{equation}
We denote by $ \mathcal{V} $  the set of all admissible control processes with respect to  equation \cref{sec:Model.00}.\\
\end{definition}
Moreover, let $ \dsl\pmb{f}:  \mathbb{R}^m \times  [0,T]  \times \Theta\rightarrow [0,\infty) $ and $ \pmb{h}: \mathbb{R}^m \rightarrow [0, \infty) $  be continuous utility functions. We define the cost function by
$$
\pmb{J}(x, t, \upsilon) =  \E_{x, t} \lf( \int^T_\zs{t} \pmb{f} (\varsigma, u, \upsilon_\zs{u}) \d u + \pmb{h} \left(\varsigma^\upsilon_\zs{T} \ri) \ri), \quad 0 \leq  t  \leq  T,
$$
where $ \E_{x, t} $ is the expectation operator conditional on $ \varsigma^{\upsilon}_\zs{t} = x$. Our goal is to solve the optimization problem \eqref{sec:Model.01++*} given by
\begin{align*}
\pmb{J}^*(x, t) : =  \sup_{\upsilon \in \mathcal{V}} \pmb{J}(x, t, \upsilon) .
\end{align*}
To this end we introduce the Hamilton function, i.e. for any $\varsigma$ and $ 0  \leq  t  \leq  T$,  with $ \pmb{q} \in \mathbb{R}^N $ and symmetric $ N \times N $  matrix $ \pmb{M} $  we set
\begin{equation}\label{eq:H}
H(\varsigma, t, \pmb{q}, \pmb{M}): =  \sup_{\theta \in \Theta} H_0(\varsigma, t, \pmb{q}, \pmb{M}, \theta ) ,
\end{equation}
where 
$$
 H_0(\varsigma, t, \pmb{q}, \pmb{M}, \theta ) : =  \check{a}'(\varsigma, t, \theta) \pmb{q} + \frac{1}{2} tr [\check{b} \check{b}'(\varsigma, t, \theta) \pmb{M}] + \pmb{f}(\varsigma, t, \theta) .
 $$
In order to find the solution to \cref{sec:Model.01++*}, we investigate the HJB equation 
\begin{align}
\begin{cases}
z_\zs{t}(\varsigma, t) + H(\varsigma, t, z_{\varsigma}(\varsigma, t), z_{\varsigma \varsigma} (\varsigma, t)) = 0, \quad t \in [0,T], \\
z(\varsigma, T) =  \pmb{h}(\varsigma), \quad \varsigma \in \mathbb{R}^N .
\end{cases}
\end{align}
Here, $ z_\zs{t} $  denotes the partial derivative of $ z $  with respect to $ t,  z_{\varsigma}(\varsigma, t) $ the gradient vector with respect to $ \varsigma $  in $ \mathbb{R}^N $  and $ z_{\varsigma \varsigma} (\varsigma, t) $  denotes the symmetric hessian matrix, that is the matrix of the second order partial derivatives with respect to $ \varsigma $.
\\
We assume the following conditions hold:
\begin{itemize}
\item[$\H_\zs{1}$)] {\sl There exists a function $ z \left( \varsigma \quad \mbox{from} \quad \C^\zs{2,1} \left( \mathbb{R}^N \times [0,T] \right) , t \right)$ from $  \bbr^N \times [0,T]  \rightarrow (0, \infty) $  which satisfies the HJB equation.}
\item[$\H_\zs{2}$)] {\sl There exists a measurable function $ \theta^*: \mathbb{R}^N \times [0,T]  \rightarrow \Theta $,   such that for all  $ \varsigma \in \mathbb{R}^N $ and $ 0  \leq  t  \leq  T $,
$$
H(\varsigma, t, z_{\varsigma}(\varsigma, t), z_{\varsigma \varsigma} (\varsigma, t)) =  H_0 (\varsigma, t, z_{\varsigma}(\varsigma, t), z_{\varsigma \varsigma} (\varsigma, t), \theta^*(\varsigma^0, t)) .
$$}
\item[$\H_\zs{3}$)]{\sl Assume that for any $\upsilon \in \cV$,  any $0 \le t \le T$ and $x$,
$$
\E_\zs{x,t} \sup_\zs{t \le u \le T} \big( z(X_\zs{u}^\upsilon, u) \big)_\zs{-} < + \infty \,.
$$
}
\item[$\H_\zs{4}$)]{\sl There exists a unique strong solution to the It\^o equation
$$
d \varsigma^*_\zs{t}  = \check{a}(\varsigma^*_\zs{t}, t) dt+ \check{b}( \varsigma^*, t) dW_\zs{t}, \quad \varsigma^*_0 = x, \quad t \geq 0  ,
$$
where $ \check{a}(., t) = \check{a}(., t,  \theta^*(., t)) $ and  $ \check{b}(., t) = \check{b}(., t, \theta^*(., t)) $. Moreover, the optimal control process $ \upsilon^*_\zs{t} =  \theta^*( \upsilon^*_\zs{t}, t) $ for $ 0  \leq  t  \leq  T $  belongs to $ \mathcal{V} $, 
and
$$
\E \sup_\zs{t \le u \le T} |z(x_\zs{u}^{*}, u)| 	<	+ \infty \,.
$$
 }
\end{itemize}
\begin{theorem}\label{thm: Ver}
Assume that conditions $ \H_\zs{1}$)- $ \H_\zs{4}$) hold
$$
\Rightarrow \upsilon_\zs{t}^* = (\upsilon_\zs{t}^*)_{0  \leq  t  \leq  T} \,,
$$
is a solution to this problem.
\end{theorem}
\proof
For $\upsilon \in \cV$, let $X^\upsilon$ be the associated wealth process with initial value $X_\zs{0}^\upsilon=x$. 
Define a stopping time
$$
\tau_\zs{n}= \inf \lf\{  s \ge t : \int^s_\zs{t}  | \check{b}'(\varsigma^{\upsilon}_\zs{u}, u)  \, \partial_\zs{\varsigma} z (\varsigma_\zs{u}^{\upsilon}, u) |^2  \d u \ge n  \ri\} \wedge T \,.
$$
Note that condition \eqref{eq: inta+b} implies that $\tau_\zs{n} \rightarrow T$ as $n \rightarrow \infty$ a.s.. 
By continuity of $z(.,.)$ and of $(\varsigma_\zs{t}^\upsilon)_\zs{0 \le t \le T}$ we obtain
\begin{equation}\label{eq:limz=h}
\lim_\zs{n \rightarrow \infty} z(\varsigma_\zs{\tau_\zs{n}}^\upsilon, \tau_\zs{n} )  =  z(\varsigma_\zs{T}^\upsilon, T)  =  \pmb{h}(\varsigma_\zs{T}^\upsilon) 
\quad
\mbox{a.s.} \,.
\end{equation}
To simplify we use the notation $\check{a}_\zs{t}= \check{a}  (\varsigma_\zs{t}, \upsilon_\zs{t}, t) $ and $ \check{b}_\zs{t}= \check{b}  (\varsigma_\zs{t}, \upsilon_\zs{t}, t)$.
Then by It\^o formula
\begin{equation} \label{eq:taylor}
\d z(\varsigma_\zs{t}, t)=z_t(\varsigma_\zs{t}, t) \d t
						+\sum_{i=1}^N \frac{\partial}{\partial\varsigma_i} z(\varsigma_\zs{t}, t)d \varsigma_{i}
						+\frac{1}{2} \sum_{i,j=1}^N \frac{\partial^2}{\partial\varsigma_i\partial\varsigma_j} z(\varsigma_\zs{t}, t) \d < \varsigma_{i}, \varsigma_\zs{j} >_\zs{k} \,.
 \end{equation}
By using the definition of $\d \varsigma$, this equation becomes
\begin{align*} 
 \d z(\varsigma_\zs{t}, t) =& z_\zs{t}(\varsigma_\zs{t}, t)
 									+( \partial z(\varsigma_\zs{t}, t))' \check{a}_\zs{t}^{\upsilon}  \d t
 									+\frac{1}{2} \mbox{tr} \big( \check{b}_\zs{t}^\upsilon (\check{b}_\zs{t}^\upsilon)'   \partial^2 z(\varsigma_\zs{t}, t)  \big)
 									 \d t 
 									 \\
 									 & + \big(\partial z (\varsigma_\zs{t}, t) \big)' \check{b}_\zs{t}^\upsilon \d W_\zs{t} \,.
\end{align*}
Taking the integration for both sides we get
\begin{align} z(\varsigma_\zs{T}, T)-z(\varsigma_\zs{t}, t)  
									=& \int_\zs{t}^T  \Big( z_\zs{u}(\varsigma_\zs{u}, u)
 									+( \partial z(\varsigma_\zs{u}, u))' \check{a}_\zs{u}^{\upsilon} \nonumber
 									+\frac{1}{2} \mbox{tr} \big( \check{b}_\zs{u}^\upsilon (\check{b}_\zs{u}^\upsilon)'   \partial^2 z(\varsigma_\zs{u}, u)  \big)
 									\\
 									&+ \int_\zs{t}^T  \big(\partial z (\varsigma_\zs{u}, u) \big)' \check{b}_\zs{t}^\upsilon  \d W_\zs{u} \,. \nonumber
\end{align} 
Add and subtract $\int_t^T \pmb{f}(\varsigma, u) \d u$ and let $z(\varsigma_\zs{T}, T)=\pmb{h}(\varsigma)$ we get
\begin{align} \label{eq: z_Ito}
z(\varsigma_\zs{t}, t)  =& \pmb{h}(\varsigma) 
									- \int_\zs{t}^T  \big(\partial z (\varsigma_\zs{u}, u) \big)' \check{b}_\zs{t}^\upsilon  \d W_\zs{u} 
 									+ \int^T_\zs{t} \pmb{f} (\varsigma_\zs{u}, u) \d u 
 									\\
									& - \int_\zs{t}^T  \Big( z_\zs{u}(\varsigma_\zs{u}, u)
 									+( \partial z(\varsigma_\zs{u}, u))' \check{a}_\zs{u}^{\upsilon} \nonumber
 									+\frac{1}{2} \mbox{tr} \big( \check{b}_\zs{u}^\upsilon (\check{b}_\zs{u}^\upsilon)'   \partial^2 z(\varsigma_\zs{u}, u)  + \pmb{f}(\varsigma_\zs{u}, u) \Big) \,.\nonumber 									
\end{align}
Take the expectation on both sides under $\varsigma$ and $t$ noting that $\E_\zs{\varsigma, t} z(\varsigma_\zs{t}, t)=z(\varsigma_\zs{t}, t)$
\begin{align*} 
z(\varsigma_\zs{t},& t)= \E_\zs{\varsigma, t} \pmb{h}(\varsigma_\zs{T}^{\upsilon})
								- \E_\zs{\varsigma, t} \int_\zs{t}^T  |  ( \check{b}(\varsigma_\zs{u},\upsilon_\zs{u}, u))' \partial z (\varsigma_\zs{u}, u) |^2 \d u 
 								    + \E_\zs{\varsigma, t}  \int^T_\zs{t} \pmb{f} (\varsigma, u) \d u 
 								    \\
						             &- \E_\zs{\varsigma, t}  \int_\zs{t}^T 
									 \Big( z_\zs{u}(\varsigma_\zs{u}, u)
 									+( \partial z(\varsigma_\zs{u}, u))' \check{a}_\zs{u}^{\upsilon} \nonumber
 									+\frac{1}{2} \mbox{tr} \big( \check{b}_\zs{u}^{\upsilon} (\check{b}_\zs{u}^\upsilon)'   \partial^2 z(\varsigma_\zs{u}, u)  + \pmb{f}(\varsigma_\zs{u}, u) \Big) \d u \,. \nonumber
\end{align*}
From the condition that 
\begin{equation}
\pmb{J}(\varsigma,\upsilon, t)= \E_\zs{\varsigma, t} \Big( \int_\zs{t}^T \pmb{f}(\varsigma_\zs{u}, \upsilon_\zs{u}, u) \d u+\pmb{h} (\varsigma_\zs{T}) \Big) \,,
\end{equation}
and 
\begin{equation}
H_0(t,\varsigma,q,M)=a'(\varsigma_t,\upsilon_\zs{t}, t) \pmb{q}+\frac{1}{2}tr[bb' (\varsigma_\zs{t},\upsilon_\zs{t}, t) \pmb{M}]+\pmb{f}(\varsigma,\upsilon, t) \,,
\label{H_o} 
\end{equation}
where $\pmb{q} =(\pmb{q}_\zs{1}, \dots, \pmb{q}_\zs{N})'  \in \bbr^N$ 
and  a symmetric  $N \times N$  matrix 
$\pmb{M}= (\pmb{M}_\zs{ij})_\zs{1 \le i,j  \le N}$, then we have
\begin{equation*}
z_\zs{t}( \varsigma, t)=\pmb{J} (\varsigma_\zs{t},\upsilon_\zs{t}, t)- \E_\zs{\varsigma, t} \left( \int_\zs{t}^T (z_\zs{u}(\varsigma_\zs{u}, u)+H_\zs{0}(\varsigma_\zs{u}, u, \pmb{q}, \pmb{M}, \upsilon)) \, \d u \right) \,.
\end{equation*}
We present the first term as 
$$
\int^{\tau_\zs{n}}_\zs{t}  \pmb{f}( \varsigma_\zs{u}, u, \upsilon_\zs{u}) \d u = \int^{\tau_\zs{n}}_\zs{t}  ( \pmb{f}( \varsigma_\zs{u}, u, \upsilon_\zs{u}))_\zs{+} \d u - \int^{\tau_\zs{n}}_\zs{t}  ( \pmb{f}( \varsigma_\zs{u}, u, \upsilon_\zs{u}))_\zs{-} \d u \,.
$$
Taking into account that 
$$
\E \int^T_\zs{0} (  \pmb{f}( \varsigma_\zs{u}, u, \upsilon_\zs{u}))_\zs{-} \d u < + \infty \,.
$$
We obtain  by the Monotone Convergence Theorem that
$$
\lim_\zs{n \rightarrow \infty}  \E \int^{\tau_\zs{n}}_\zs{0} \pmb{f}( \varsigma_\zs{u}, u, \upsilon_\zs{u}) \d u  = 
\E \int_\zs{t}^T  \pmb{f}( \varsigma_\zs{u}, u, \upsilon_\zs{u}) \d u \,.
$$
From the following conditions:
\begin{equation}
z_\zs{t}(\varsigma, t)+ H(\varsigma,t, \pmb{q}, \pmb{M})=0 \,,
\end{equation}
and the Hamilton function
\begin{equation}
H(\varsigma,t, \pmb{q}, \pmb{M})=\sup_\zs{\substack{\upsilon \in \Theta}} H_0(\varsigma, t, \pmb{q}, \pmb{M} , \upsilon) \,.
\end{equation}
Then $z(.,.)$ becomes
\begin{align*}
z(\varsigma_\zs{t}, t)
			=\pmb{J} (\varsigma_\zs{t}, \upsilon_\zs{t}, t)+ \E_\zs{\varsigma, t} \left( \int_\zs{t}^T \left( H(\varsigma, u, \pmb{q}, \pmb{M})-H_0(\varsigma, u, \pmb{q}, \pmb{M} ,\upsilon) \right) \d u \right) \,.
\end{align*}
Moreover, taking into account that 
$$
\E_\zs{\varsigma, t} \sup_\zs{n \ge 1} (z( \varsigma_\zs{\tau_\zs{n}}^\upsilon , \tau_\zs{n} ))_\zs{-} \le 
\E_\zs{\varsigma, t} \sup_\zs{ 0 \le t \le T} (z( \varsigma_\zs{t}^\upsilon , t ))_\zs{-} \le + \infty \,.
$$
We thereby  Fatou's Lemma obtain that 
$$
\lim_\zs{n \rightarrow \infty} \E_\zs{\varsigma, t} z( \varsigma_\zs{\tau_\zs{n}}^\upsilon, \tau_\zs{n}) \ge 
 \E_\zs{\varsigma, t} \lim_\zs{n \rightarrow \infty} z(\varsigma_\zs{\tau_\zs{n}}^\upsilon, \tau_\zs{n})  =
  \E_\zs{\varsigma, t}  z( \varsigma_\zs{T}^\upsilon, T) =
   \E_\zs{\varsigma, t}  \pmb{h}( \varsigma_\zs{T}) \,.
$$
Finally, we obtain that
\begin{align*}
z(\varsigma, t) \ge 
				\E_\zs{\varsigma, t} \lf( \int_\zs{t}^T \pmb{f}(\varsigma_\zs{u}^\upsilon, u, \upsilon_\zs{u}) \d u +  \pmb{h}( \varsigma_\zs{T} ) \ri)
			= J(\varsigma, t, \upsilon) \,.
\end{align*}
$$
z(\varsigma, t) \ge \pmb{J} (\varsigma, \upsilon, t) \,.
$$
Therefore, $z( \varsigma, t) \ge J^*( \varsigma, t)$ for all $0 \le t \le T$.
Similarly, replacing $\upsilon$ in \eqref{eq: z_Ito} by $\upsilon^*$ as defined by $H_\zs{2}- H_\zs{3}$ we obtain
$$
z(\varsigma, t)= \E_\zs{ \varsigma, t} \int_\zs{t}^{\tau_\zs{n}} \pmb{f}( \varsigma_\zs{u}^\upsilon, u, \upsilon_\zs{u}) \d u +  \E_\zs{ \varsigma, t} z( \varsigma_\zs{\tau_\zs{n}}^*,  \tau_\zs{n}) \,.
$$
Condition $H_\zs{4}$ implies that the sequence $(z( \varsigma_\zs{\tau_\zs{n}}^*,  \tau_\zs{n}))_\zs{n \in \bbn}$ is uniformly integrable. 
Therefore, by \eqref{eq:limz=h}
$$
\lim_\zs{n \rightarrow \infty}  \E_\zs{ \varsigma, t} z( \varsigma_\zs{\tau_\zs{n}}^*, \tau_\zs{n} ) 
				= \E_\zs{ \varsigma, t} \lim_\zs{n \rightarrow \infty} z( \varsigma_\zs{\tau_\zs{n}}^*,  \tau_\zs{n}) 
				= \E z( \varsigma_\zs{T}, T)
				= \E_\zs{ \varsigma, t}  \pmb{h}( \varsigma_\zs{T}^*) \,,
$$
and we obtain 
\begin{align*}
z( \varsigma, t)&=  \lim_\zs{n \rightarrow \infty}  \E_\zs{ \varsigma, t}   \int_\zs{t}^{\tau_\zs{n}} \pmb{f}( \varsigma_\zs{u}^*, u, \upsilon_\zs{u}^*) \d u
			+  \lim_\zs{n \rightarrow \infty} \E_\zs{ \varsigma, t}  z( \varsigma_\zs{\tau_\zs{n}}^*,  \tau_\zs{n}) 
			\\
			&= \E_\zs{ \varsigma, t} \lf( \int_\zs{t}^T \pmb{f}( \varsigma_\zs{u}^*, u, \upsilon_\zs{u}^*) \d u +  \pmb{h}( \varsigma_\zs{T}^* ) \ri)
			= J( \varsigma, t, \upsilon^*) \,.
\end{align*}
We arrive at $z(\varsigma, t)= J^*(\varsigma, t)$. 
This proves \cref{thm: Ver}. 
\endproof
%
\begin{remark}
The difference in \cref{thm: Ver} from the verification theorem from \cite{BerdjanePergamenchtchikov2013} is that the functions 
$\pmb{f}$
and 
$\pmb{h}$
are positive but from the  logarithmic utilities these functions are negative.
So, we can not use directly the verification theorem in \cite{BerdjanePergamenchtchikov2013}.
\end{remark}

\newpage
\section{Proofs}

\subsection{Proof of \cref{thm:solHJB}}
Now, by taking the derivatives of $z(\varsigma, t)$ defined in \eqref{eq: z} with respect to $t$ and $s$ and apply them into  equation \eqref{sec:HJB.4} we obtain
\begin{align*}
 s' \dot g(t) s &+  \dot f(t) + r \rho(t) +  \sum_\zs{k,i=1}^d ( <\sigma \sigma'>_\zs{ki} (g_\zs{ki} + g_\zs{ik}))  - \ln \rho(t) -1 
\\
& +   \sum_\zs{j=1}^d   \sum_\zs{l=1}^d A_\zs{jl} s_\zs{l} < (g + g' ) s >_\zs{j}
 + \frac{ \rho(t) \wh{s}'  (\sigma \sigma')^{-1} \wh{s} }{2}  =0 \,,
 \end{align*}
 where the dot $"\cdot " $ denotes the first derivative and $g$ is  a $d \times d$ matrix  defined in \eqref{eq: gmatrix}.
Then this can be written as 
\begin{align*}
s' \Big(  \dot g(t) & + \frac{1}{2} \rho(t) A_\zs{1}' (\sigma \sigma')^{-1} A_\zs{1} - A' (g + g') \Big) s 
+\dot f(t) + \sum_\zs{k,i=1}^d <\sigma \sigma'>_\zs{ki} (g_\zs{ki} + g_\zs{ik})) -1
\\
& - \ln \rho(t) + r \rho(t)  =0 \,.
\end{align*}
After calculation we get that for $s \in \bbr^d$,
$$
 f(t) =  \sum^d_\zs{k,i=1} <\sigma \sigma' >_\zs{ki}( \wt g_\zs{ki}(v) + \wt g_\zs{ik}(v)) +f_\zs{0}(t) \,,
$$
and
$$
s' \big(  \dot g(t) + \frac{1}{2} \rho(t) A_\zs{1}' (\sigma \sigma')^{-1} A_\zs{1} - A' (g + g') \big) s =0, \quad g(T)=0 \,,
$$
where
$$
\wt g(t)=  \int^T_\zs{t}  g(v) \d v
\quad
\mbox{and}
\quad
f_\zs{0}(t)= \frac{1}{2} r \Big( t^2- 2t(T+1) +T(T+2)  \Big) + \rho(t) \ln \rho(t) \,.
$$
The last term in the preceding equation can be written as
\begin{align*}
<A' (g +g')>_\zs{ij} &= \sum_\zs{l=1}^d   <A'>_\zs{il} (g_\zs{lj} + g_\zs{jl}) \,,
							\\
							&=  \sum_\zs{l=1}^d   \lf(  <A>_\zs{li} g_\zs{lj} + <A>_\zs{li} g_\zs{jl} \ri) \,.
\end{align*}
Let we denote  by $H= (h_\zs{ij})_\zs{1 \le i,j \le d}$,  where $h= \mbox{vect(H)}$ a vector in $\bbr^m$  such that 
$h= (h_\zs{1}, h_\zs{2}, \dots, h_\zs{m})$, with 
$h_\zs{(j-1)d+ i} = <H>_\zs{i,j}$ and 
$Z(t)= \mbox{vect}(g(t))$ where $Z_\zs{(j-1)d+i} = g_\zs{ij}$.
Therefore, the last equation becomes
\begin{align*}
<A' (g +g')>_\zs{ij}  &=  \sum_\zs{l=1}^d  <A>_\zs{li}    \lf(   Z_\zs{(j-1)d + l}+  Z_\zs{(l-1)d+j} \ri) \,,
							\\
							& =   \sum_\zs{l=1}^d  <A>_\zs{li}    Z_\zs{(j-1)d + l} +  \sum_\zs{l=1}^d  <A>_\zs{li}     Z_\zs{(l-1)d+j} \,,
							\\
							&= \sum_\zs{l=1}^d  \sum_\zs{k=1}^d  \lf( <A>_\zs{li} \pmb{1}_\zs{\{k=j \}}   Z_\zs{(k-1)d + l}
							+    <A>_\zs{k i} \pmb{1}_\zs{\{l=j \}}   Z_\zs{(k -1)d + l } \ri) \,.
\end{align*}
This can be written in the following form 
$$
\mbox{Vect} (A' (g +g') )   = \Gamma Z \,,  
$$
where $\Gamma= (\check{\gamma}_\zs{s,t})$ and $  \wh{\gamma}_\zs{s,t} =  <A>_\zs{li} \pmb{1}_\zs{\{k=j \}} + <A>_\zs{k i} \pmb{1}_\zs{\{l=j \}} $, with  
$s= (j-1)d +i$ 
and 
$t=(k-1)d +l$.
Therefore, for all $m \times m$ matrix $\Gamma $, with $m=d^2$,
$$
<A' (g +g')>_\zs{ij}  =  <\Gamma Z>_\zs{ij} \,,
$$
where $\Gamma = (\wh{\gamma}_\zs{s,t})_\zs{1 \le s, t \le m}$.
Thus equation \eqref{eq: gmatrix} can be written as
$$
\dot Z - \Gamma Z + \frac{1}{2} \rho(t) \wt{b}=0 \,, 
\quad
Z(T)=0 \,,
$$
where $\wt{b}= \mbox{vect}(A_\zs{1}' (\sigma \sigma')^{-1} A_\zs{1} ) \in \bbr^{m}$.
Therefore, the solution of $Z(t)$ is given by
\begin{equation}
Z(t)= \wt{b} \frac{1 }{ 2} \int_\zs{t}^T \rho(v) \me^{ \Gamma (v-t)} \d v \,.
\end{equation}
This proves \cref{thm:solHJB}. 
\fdem

\subsection{Proof of \cref{thm:strgy}}
We apply the Verification \cref{thm: Ver} to Problem \eqref{sec:Model.01++*} for the stochastic control differential equation \eqref{sec:Md.3}. 
For fixed $\u=(\alpha, c)$, where $\alpha \in \bbr^d$ and $c \in [0, \infty)$, the coefficients in model \eqref{sec:Model.00} are defined as 
\begin{align*}
\check{a}(\varsigma, \u) = \begin{pmatrix} 
\begin{array}{c} r x- \alpha' \wh{s}	 - c \\
											A s
											\end{array}
											\end{pmatrix} ,
\quad
\check{b}(\varsigma, \u ) =\begin{pmatrix} 
\begin{array}{c}   \alpha'  \sigma \\
 \sigma  
   \end{array} \end{pmatrix}  ,
\quad
\mbox{and}
\quad \pmb{h}(x)= \ln x \,.
\end{align*}
This implies immediately condition $H_\zs{1})$. 
Moreover, by Definition \eqref{def:1}, the coefficients are continuous, hence \eqref{eq: inta+b} holds for every $\u \in \cV$. 
To check $H_\zs{1}) - H_\zs{3})$ we calculate the Hamilton function \eqref{eq: inta+b} for Problem \eqref{sec:Model.01++*}. 
We have
$$
 H(\varsigma, \pmb{q}, \pmb{M})= \sup_\zs{\u \in \bbr^d \times \bbr_\zs{+}}   H_\zs{0}(\varsigma, \pmb{q}, \pmb{M}, \u) \,,
$$
where
$$
H_0(\varsigma, \pmb{q}, \pmb{M}, \u)=\check{a}'(t, \varsigma, \u) \pmb{q} + \frac{1}{2} \pmb{tr} [\check{b} \check{b}'(t, \varsigma, \u) \pmb{M}]+U(c) \,.
$$
As
$$
<\check{b} \check{b}'>_\zs{1+k, 1+j} = \sum_\zs{l=1}^m \check{b}_\zs{1+k, l} \check{b}_\zs{1+j, l} = \sum_\zs{l=1}^m \sigma_\zs{kl} \sigma_\zs{jl} = <\sigma \sigma'>_\zs{kj} \,,
$$
and
$$
\mbox{tr} \lf( \check{b} \check{b}' \pmb{M} \ri) = \alpha' \sigma \sigma' \alpha \pmb{M}_\zs{11}  
																+ 2 \sum_\zs{j=1}^d <\sigma \sigma' \alpha>_\zs{j} \pmb{M}_\zs{1, 1+j}
															     + \sum_\zs{k,j=1}^d  <\sigma \sigma'>_\zs{kj} \pmb{M}_\zs{1+k, 1+j} \,.
$$
Therefore, $H_\zs{0}$ can be written as
$$
H_\zs{0} (\varsigma, \pmb{q}, \pmb{M}, \u)=  rx  \pmb{q}_\zs{1}
																	+ \sum_\zs{j=1}^d \wt{s}_\zs{j} \pmb{q}_\zs{1+j}
																	 + \sum_\zs{k, j=1}^d <\sigma \sigma'>_\zs{kj}  \pmb{M}_\zs{1+k, 1+j}
																	- c  \pmb{q}_\zs{1} + \ln c + J(\alpha) \,,
$$
where 
$$
J(\alpha)=   \frac{\alpha' \sigma \sigma' \alpha }{2} \pmb{M}_\zs{11} 
				  +  \sum_\zs{j=1}^d <\sigma \sigma' \alpha>_\zs{j} \pmb{M}_\zs{1, 1+j}
																	- \alpha' \wh{s}  \pmb{q}_\zs{1} \,.
$$
Now in order to find the Hamilton function we have to maximize $H_\zs{0}$,
$$
\max_\zs{\alpha} J(\alpha) = \frac{ \tau' (\sigma \sigma')^{-1} \tau}{2 |\pmb{M}_\zs{11}|} \,,
$$
where $\tau= \sigma \sigma' \mu -  \pmb{q}_\zs{1} \wh{s} $ and $\mu= (\pmb{M}_\zs{1, 1+1}, \dots, \pmb{M}_\zs{1, 1+d})' $. 
Therefore, 
$$
H(\varsigma, \pmb{q}, \pmb{M}) = r x \pmb{q}_\zs{1} 
													+ \sum_\zs{j=1}^d \wt{s}_\zs{j} \pmb{q}_\zs{1+j}
													+ \sum_\zs{k, j=1}^d <\sigma \sigma'>_\zs{kj} \pmb{M}_\zs{1+j, 1+k}
													-1
													-\ln \pmb{q}_\zs{1}
													+\frac{ \tau' (\sigma \sigma')^{-1} \tau}{2 |\pmb{M}_\zs{11}|} \,.
$$
Therefore, the HJB equation can be written as 
\begin{align*}
z_\zs{t} +r x z_\zs{x}
			+ \sum_\zs{j=1}^d \wt{s}_\zs{j} z_\zs{s_\zs{j}}
              + \sum_\zs{k, j=1}^d <\sigma \sigma'>_\zs{kj} z_\zs{s_\zs{j} s_\zs{k}}
													-1
													-\ln z_\zs{x}
													+\frac{ \tau' (\sigma \sigma')^{-1} \tau}{2 |z_\zs{xx}|} =0 \,,
\end{align*}
where $\tau= \tau(\varsigma, t)$ and $\tau_\zs{j}= \sum_\zs{l=1}^d <\sigma \sigma'>_\zs{il} z_\zs{x s_\zs{l}}- z_\zs{x} \wh{s}_\zs{j}$.
By taking 
$$
z(x, s, t)= (T-t+1) \ln x + s' g(t) s + f(t) \,,
$$
with $g$ and $f$  are given in \eqref{eq: gmatrix}, 
we obtain
$$
s' \dot g(t) s + \dot f(t) 
		 + r \rho(t)
		 + s' A'  ( g+  g' ) s
		 + \mbox{tr} ( \sigma \sigma'  (g +g') )
		 -1
		 - \ln \rho(t)
		 +\frac{ \rho(t) \wh{s}'  (\sigma \sigma')^{-1} \wh{s} }{2} =0 \,.
$$
So, $z(\varsigma, t)$ given in \eqref{eq: z} is the solution to the HJB equaiton \eqref{Hamilton--Jacobi--Bellman equation}. 
One can check directly that the strategy $\nu$ in  with optimal strategy in \eqref{opt. stgy.11} satisfies the  conditions $H_\zs{1})- H_\zs{3})$. Now to check condition $H_\zs{4})$ we have to verify that
$$
\sup_{0 \le t \le T} \E_{ \varsigma} |  z( \varsigma^*_{t}, t)| < + \infty \,.
$$
Thus,  as 
$$
 z(\varsigma, t)= (T-t+1) \ln x + s' g(t) s + f(t) \,,
 $$
  and as $g(t)$ and $f(t)$ are bounded functions and  
$X^*_\zs{t}$ is given by 
$$
X^*_\zs{t}= x \exp \bigg\{ \int^t_\zs{0} \lf( \check{a}^*(u)- \left(\check{b}^*(u) \right)^2/2 \right) \d u+  \int^t_\zs{0} \check{b}^*(u) \d W_\zs{u}   \bigg\}   \,,
$$
 then we have to show that 
$$
 \sup_\zs{\tau \in \cM_\zs{t}} \E \lf( \lf( | \ln (X^*_\zs{\tau}) | + S_\zs{\tau}^2 \ri)
  \Big|  X_\zs{t}= x, S_\zs{t}= s \ri) < + \infty  \,.
 $$
Moreover, note that
 $$
 S_\zs{\tau}= \me^{- \varkappa (T-t)}  s+ \xi_\zs{t, \tau} 
 \quad
 \mbox{and}
 \quad
  \xi_\zs{t, \tau} =  \sigma  \me^{- \varkappa \tau} \int^\tau_\zs{t} \me^{ \varkappa u} \d W_\zs{u} .
 $$
Since
$|S_\zs{\tau}| \le |s|+ |\xi_\zs{t, \tau}|$,
  one needs to check that 
  $$
 \sup_\zs{0 \le t \le T} \E \big( \ln(X^*_\zs{t}) +   \xi^2_\zs{t}  \big)   < + \infty \,.
 $$
From equation \eqref{eq:dXt*}, we have that 
$|\check{a}^*(t)| \le c_\zs{1} (1+ S^2_\zs{t})$ and $|\check{b}^*(t)| \le c_\zs{2} |S_\zs{t}|$.
Thus by the OU process, 
$\E \int^T_\zs{0} |\check{b}^*(t)|^m  \d u    \le    +\infty $,
which implies that
$$
\E  \lf( \int^T_\zs{0}  \check{b}^*(u) \d W_\zs{u}  \ri)^2 
						=
						\E \int^T_\zs{0} |\check{b}^*(u)|^2  \d u 
						\le 
						c \,.
$$
Therefore, 
$$
\E | \ln X^*_\zs{t} |
							 \le  
							 \E  \int^T_\zs{0}  \lf(  |\check{a}^*(u)| + \frac{1}{2} \lf( \check{b}^*(u) \ri)^2  \ri)  \d u 
							+ \sqrt{  \E \lf( \int^T_\zs{0}  \check{b}^*(u) \d u \ri)^2  }  
							< + \infty \,.  
$$
This proves \cref{thm:strgy}. 
\endproof
\fdem

\section{Appendix}\label{sec:A}
The R simulation codes:
\begin{verbatim}
## defining the variables

T=1
r=0.01
kappa=0.1
kappa1=kappa+r
sigma=0.5
x=100
s<-seq(-10,10, length=30)
t<-seq(0,1, by=.1)

## The function g
g<- function(t) (-kappa1^2/2*sigma^2 )*((-2*kappa*exp(2*kappa*(t-T)) 
                           + t-T-1 + exp(2*kappa*(t-T)) - 1 )/(4* kappa^2))

## The function f
f<-function(t) sigma^2*g(t)  + T-t-(T-t+1)* log(T-t+1)

## The function Z
Zvarsigma<-function(s,t) log(T-t+1) +s^2*g(t)+f(t)

## The plot  ###

z<-outer(s,t, Zvarsigma)
jet.colors <- colorRampPalette( c("blue", "green") )
nbcol <- 100
color <- jet.colors(nbcol)
nrz <- nrow(z)
ncz <- ncol(z)
zfacet <- z[-1, -1] + z[-1, -ncz] + z[-nrz, -1] + z[-nrz, -ncz]
facetcol <- cut(zfacet, nbcol)

persp(s, t, z, col = color[facetcol], phi = 30, theta = -30, xlab = "S", 
            ylab = "t", zlab ="Z", ticktype = "detailed")
# ticktype -- to give details in the numbers or values of each variable 

## The strategies ##
rho<-T-t+1
astar<- r+(kappa1*s^2/sigma^2)-1/rho
bstar<--kappa1*s/sigma
xstar<-seq(0.1,100)
alphastar<- s*xstar/sigma^2
persp(s, xstar,alphastar, xlab = "S", ylab = "xstar", zlab ="alpha^*")
cstar<-xstar/(T-t+1)
persp(xstar, t,  cstar, xlab = "xstar", ylab = "t", zlab ="c^*")
\end{verbatim}

\newpage

\end{document}